\documentclass[aps,prl,twocolumn,superscriptaddress,longbibliography]{revtex4-1}
\usepackage{amsmath}	
\usepackage{graphicx}	
\usepackage{color}
\usepackage{gensymb}
\usepackage{hyperref}
\usepackage{braket}
\usepackage{upgreek}
\usepackage{bm}

\begin{document}
	
\title{The valley Zeeman effect in inter- and intra-valley trions in monolayer WSe$_2$}
	
\author{T. P. Lyons}
\email{tplyons1@sheffield.ac.uk}
\author{S. Dufferwiel}
\affiliation{Department of Physics and Astronomy, University of Sheffield, Sheffield S3 7RH, UK}
\author{M. Brooks}
\affiliation{Department of Physics, University of Konstanz, D-78464, Konstanz, Germany}
\author{F. Withers}
\affiliation{School of Physics and Astronomy, University of Manchester, Manchester M13 9PL, UK}
\affiliation{Centre for Graphene Science, CEMPS, University of Exeter, Exeter, EX4 4QF, UK}
\author{T. Taniguchi}
\affiliation{National Institute for Materials Science, Tsukuba, Ibaraki 305-0044, Japan}
\author{K. Watanabe}
\affiliation{National Institute for Materials Science, Tsukuba, Ibaraki 305-0044, Japan}
\author{K. S. Novoselov}
\affiliation{School of Physics and Astronomy, University of Manchester, Manchester M13 9PL, UK}
\author{G. Burkard}
\affiliation{Department of Physics, University of Konstanz, D-78464, Konstanz, Germany}
\author{A. I. Tartakovskii}
\email{a.tartakovskii@sheffield.ac.uk}
\affiliation{Department of Physics and Astronomy, University of Sheffield, Sheffield S3 7RH, UK}
	
\date{\today}
	
\begin{abstract}
	Monolayer transition metal dichalcogenides (TMDs) hold great promise for future information processing applications utilizing a combination of electron spin and valley pseudospin. This unique spin system has led to observation of the valley Zeeman effect in neutral and charged excitonic resonances under applied magnetic fields. However, reported values of the trion valley Zeeman splitting remain highly inconsistent across studies. Here, we utilize high quality hBN encapsulated monolayer WSe$_2$ to enable simultaneous measurement of both intervalley and intravalley trion photoluminescence. We find the valley Zeeman splitting of each trion state to be describable only by a combination of three distinct g-factors, one arising from the exciton-like valley Zeeman effect, the other two, trion specific, g-factors associated with recoil of the excess electron. This complex picture goes significantly beyond the valley Zeeman effect reported for neutral excitons, and eliminates the ambiguity surrounding the magneto-optical response of trions in tungsten based TMD monolayers.
\end{abstract}
	
	
\maketitle
	
Over the past several years, optical investigations of monolayer transition metal dichalcogenides (TMDs) have generated significant scientific interest \cite{wang2018colloquium}. These layered semiconductors show remarkable properties when reduced to a single atomic layer, such as an indirect to direct band gap transition \cite{mak2010atomically, splendiani2010emerging}, alongside a regime of coupled spin and valley physics \cite{yu2014valley, xu2014spin, mak2012control}. Low temperature photoluminescence (PL) of monolayer TMDs, such as tungsten diselenide (WSe$_2$), exhibits spectra dominated by excitonic emission in the near infra-red, where a range of biexcitonic and trionic complexes have been reported \cite{jones2013optical, wang2014valley, mak2013tightly, you2015observation}.

The valley degree of freedom exhibited by monolayer WSe$_2$ and other TMDs, which allows excitons to occupy degenerate but momentum-opposite states within the Brillouin zone, opens prospects for information encoding and processing exploiting the valley pseudospin \cite{schaibley2016valleytronics}. In WSe$_2$ carriers may occupy either the +K or -K valleys, where they are robust against intervalley scattering due to the large momentum transfer needed to cross the Brillouin zone, and the energy transfer required to overcome the large spin-orbit splitting at the conduction and valence band edges, which is opposite in the two valleys by time reversal symmetry \cite{xiao2012coupled}. A further property of monolayer TMDs is the locking of the polarization of optically bright transitions to the valley pseudospin: electron hole pairs in the +K (-K) valley are coupled to $\sigma^+$ ($\sigma^-$) polarized light. This allows optical generation and addressability of valley polarized excitons \cite{mak2012control}, along with their more elaborate complexes such as charged excitons (trions) \cite{singh2016long} and exciton-polaritons \cite{dufferwiel2017valley}.

A recently observed consequence of the coupled spin and valley regime inherent to monolayer WSe$_2$ is the valley Zeeman effect \cite{aivazian2015magnetic, srivastava2015valley, wang2015magneto}. Here, an external magnetic field normal to the monolayer lifts the degeneracy between valley polarized states, such that excitonic resonances in the +K and -K valleys will shift spectrally away from one another. It has been reported that this energy splitting depends on two different magnetic moments, an intracellular contribution, arising from the tungsten d-orbitals in the valence band, and an intercellular contribution from finite Berry curvature at the +K and -K points \cite{aivazian2015magnetic, srivastava2015valley, wang2015magneto}. Direct optical measurement of the valley Zeeman splitting is possible thanks to the locking of light helicity to the valley pseudospin, and allows extraction of a valley Zeeman g-factor for a given spectral resonance. While the valley Zeeman splitting of the neutral exciton is fairly consistently reported to be $E(\sigma^+) - E(\sigma^-)\approx -4 \mu_B B$ \cite{aivazian2015magnetic, srivastava2015valley, wang2015magneto}, where $\mu_B$ is the Bohr magneton, values reported for the negatively charged trion vary from $-4\mu_B B$ to $-13\mu_B B$ \cite{wang2015magneto, srivastava2015valley, koperski2017optical, koperski2015single}, and are the subject of some speculation and ambiguity as to the cause of the reported variation.

\begin{figure*}
	\center
	\includegraphics{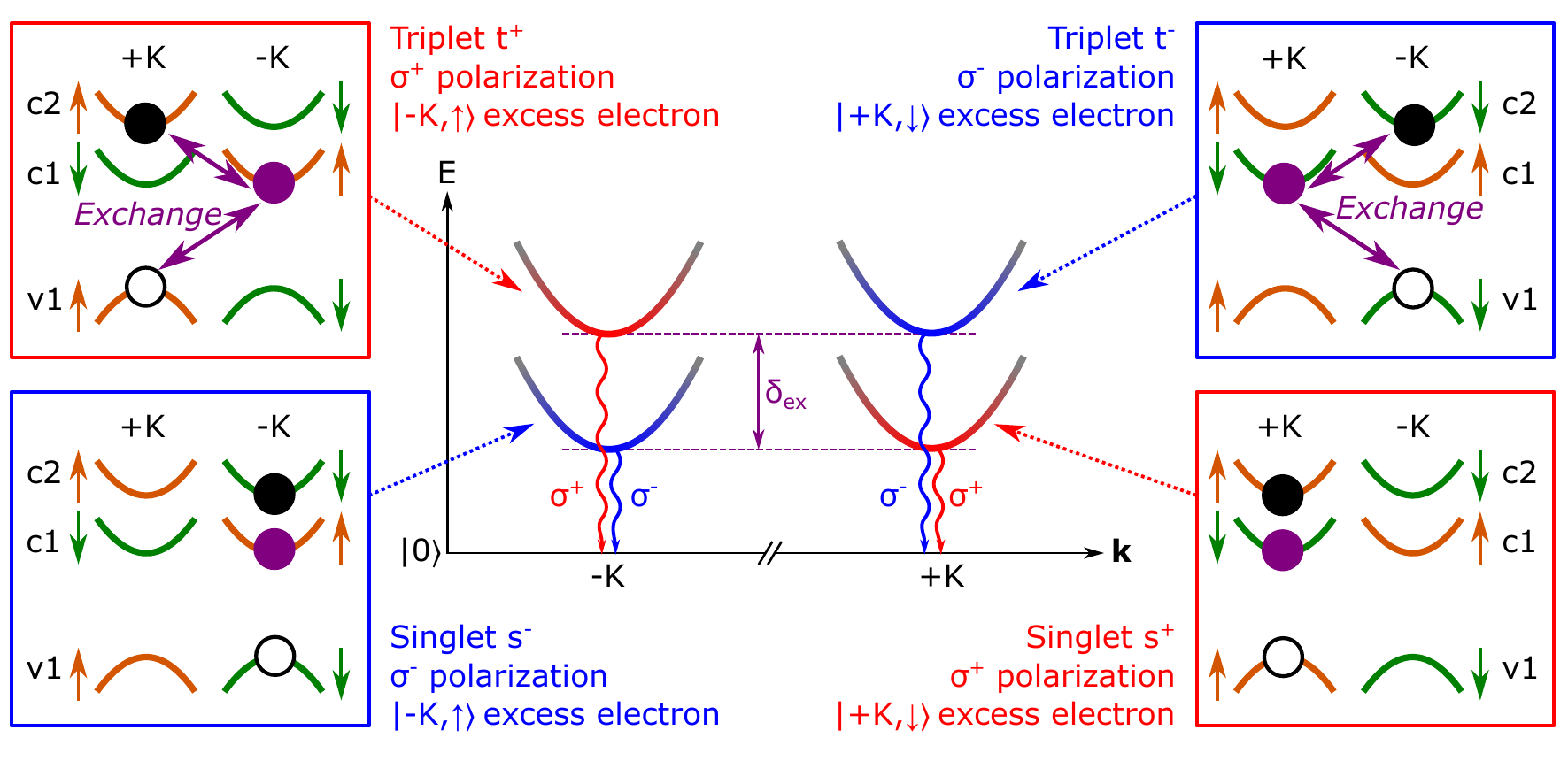}
	\caption{\label{fig:figure1} Generalized dispersion relations $E(\textbf{k})$, where $\textbf{k}$ is the centre of mass wavevector, of the four optically bright ground state negative trions in monolayer WSe$_2$. Side panels illustrate the spin-valley configurations of the constituent carriers of each trion variety. v1 is the topmost spin-subband within the valence band, while $c1$ ($c2$) is the lower (upper) energy spin-orbit split conduction band. Based on optical selection rules, the black electron and hole are the recombining pair, while the purple electron is excess and occupies band $c1$. Orange (green) conduction and valence band states are spin up (down). Red (blue) colours denote $\sigma^+$ ($\sigma^-$) helicity of the bright transition. The trion states are labelled $X^Y$ where $X = s, t$ for $singlet$ or $triplet$, $Y = +, -$ for $\sigma^+$ or $\sigma^-$ emission helicity. Purple arrows indicate the intervalley Coulomb exchange interaction, present only in triplet trions, which raises the energy of the triplets by an amount $\delta_{ex}$ above the singlets. This energy gap gives rise to the trion fine structure in emission.}
\end{figure*}

In monolayer WSe$_2$, optical selection rules dictate that negative trions must have an electron with the same spin and valley index as the hole in order to allow radiative recombination. As such, an electron must always occupy the upper spin state of the conduction band ($c2$ in Fig.~\ref{fig:figure1}), allowing the excess electron to occupy the lower energy conduction band spin state, in either valley ($c1$ in Fig.~\ref{fig:figure1}). This gives a total of four different ground state bright A-trion configurations, which are illustrated in Fig.~\ref{fig:figure1}. Two of these trion configurations are ``intravalley", with all three carriers in the same valley, and the other two are ``intervalley", with the excess electron in the opposite valley to the e-h pair which may recombine with the emission of a photon \cite{aivazian2015magnetic, singh2016long}. It is convenient to define these trion configurations as \emph{singlet} and \emph{triplet} trions, respectively, where the cumulative spin of the electron pair determines the classification. As a result we can define the four ground state trion configurations as $s^{+}, s^{-}, t^{+}, t^{-},$ where \emph{s} and \emph{t} denote singlet and triplet, and $+$ and $-$ denote the circular polarization of the optically bright transition of the state. For clarity these are labelled in Fig.~\ref{fig:figure1}.

The intervalley Coulomb exchange interaction between the e-h pair and the excess electron in triplet trions raises their energy relative to the singlet by an amount $\delta_{ex}$, expected to be a few meV \cite{yu2014dirac}. In luminescence, this energy gap gives rise to trion fine structure \cite{jones2016excitonic, singh2016long}, as depicted schematically in Fig.~\ref{fig:figure1} for the case of zero external magnetic field. In monolayer WS$_2$, similar trion fine structure has been observed \cite{plechinger2016trion, vaclavkova2018singlet}, and magneto-optical measurements have uncovered inequivalent valley Zeeman g-factors for the two fine structure components \cite{plechinger2016excitonic}. However, no thorough explanation has been given for this difference. Furthermore, to our knowledge no detailed magneto-optical study of the WSe$_2$ trion fine structure has yet been reported. It is highly likely that the complex nature of WSe$_2$ trions, having four distinct valley configurations, is the root cause of the wide ranging and apparently random valley Zeeman g-factors so far measured.

In this work, we report that the relative intensity of singlet and triplet trions has a strong temperature dependence, such that heating a WSe$_2$ monolayer from 4 K to 30 K thermally populates the triplet states allowing simultaneous measurement of the magneto-optical response of the trion fine structure components. We observe that different trion valley configurations have inequivalent rates of field-dependent spectral shift, which is incompatible with the valley Zeeman interpretation reported for neutral excitons. We extract a true valley Zeeman splitting of $(-8.3\pm0.2)\mu_B B$ for all trion states, which may be measured only by consideration of both singlet and triplet trions, and attribute its discrepancy from the known atomic orbital contribution of $\sim -4\mu_B B$ to a strong Berry curvature associated magnetic moment. However, we observe this true trion valley Zeeman splitting to be masked by energetic recoil processes of the additional electron, which modify the measured trion energy shifts in low temperature magneto-PL studies and are likely to depend heavily on external factors such as doping level, which vary from sample to sample. This work removes the ambiguity surrounding the variation of trion valley Zeeman splittings reported in literature, by revealing the interplay between different trion complexes and external magnetic fields.

\section{Results}

\subsection{Temperature dependent trion photoluminescence}

\begin{figure}
	\center
	\includegraphics{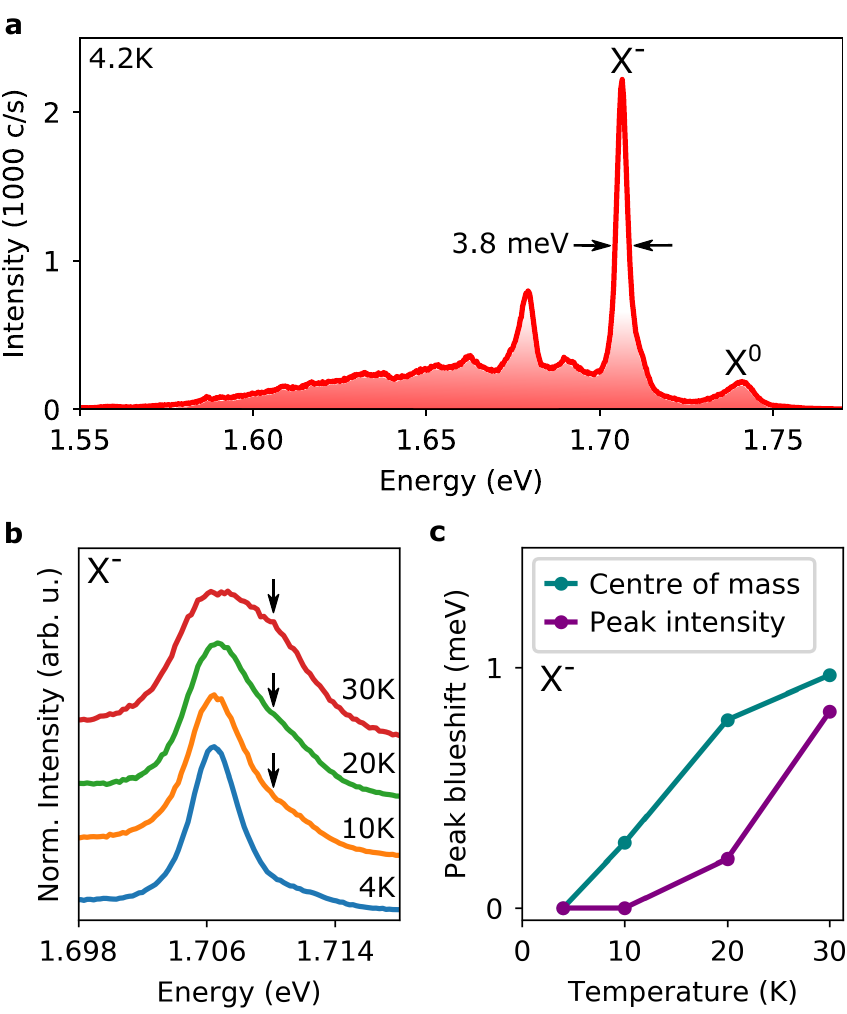}
	\caption{\label{fig:figure2} (a)  Photoluminescence spectrum from the sample under non-resonant laser excitation at 4.2 K. Peaks attributed to the neutral A-exciton (X$^0$) and trion (X$^-$) can be seen. The trion linewidth is 3.8 meV. (b) Temperature dependence of the trion photoluminescence from 4.2 K to 30 K. Increasing thermal population of the higher energy triplet states can be inferred from the increasing spectral weight of the high energy shoulder of the trion peak, indicated by the black arrows. (c) Temperature dependent blueshift of the trion emission feature, extracted using two different methods (see main text for definitions). The blueshift discrepancy between the two methods strongly indicates temperature dependent fine structure.}
\end{figure}

The sample used in this investigation consists of a monolayer of mechanically exfoliated WSe$_2$, encapsulated on both sides by few-layer hexagonal boron nitride (hBN) \cite{taniguchi2007synthesis}. Encapsulation in this manner is known to be responsible for narrow excitonic PL linewidths in TMD monolayers \cite{cadiz2017excitonic}, and is here responsible for a trion linewidth of 3.8 meV at 4.2 K, significantly narrower than the typical 10 - 20 meV values for bare WSe$_2$ \cite{jones2013optical} and approaching the intrinsic homogeneous linewidth \cite{singh2015intrinsic}. Polarization resolved PL spectra from the sample at 4 K can be seen in Fig.~\ref{fig:figure2}a. Peaks corresponding to the neutral A exciton ($X^0$) and negatively charged trion ($X^-$) are visible, as is a lower energy band of localised emission, typical of WSe$_2$ \cite{jones2013optical}.

We observe in Fig.~\ref{fig:figure2}b, upon heating the sample from its base temperature of 4.2 K, an increasing spectral weight of the high energy shoulder of the trion peak, which we attribute to increasing thermal population of the triplet states. At 30 K, the triplet emission becomes of comparable intensity to the singlet. Analysis of the peak position of trion emission by two different methods confirms the presence of fine structure, as shown in Fig.~\ref{fig:figure2}c. The centre of mass method is calculated as $(\sum (E \times I)_n)/(\sum I_n)$, where $E$ and $I$ are the energy and intensity of the $n$th pixel, whereas the ``peak intensity" is the energy of the pixel with the most counts per second. If trion emission is a single peak with no fine structure, then these two methods should show roughly the same behaviour. The fact that they do not agree confirms the presence of fine structure components with relative intensities dependent on temperature. For the remainder of this investigation, the sample was maintained at 30 K, where the comparable populations of singlet and triplet states allows the greatest insight into the magneto-optical response of the trion fine structure.

\subsection{Magneto-optical response of singlet and triplet trions in WSe$_2$}

\begin{figure*}
	\center
	\includegraphics{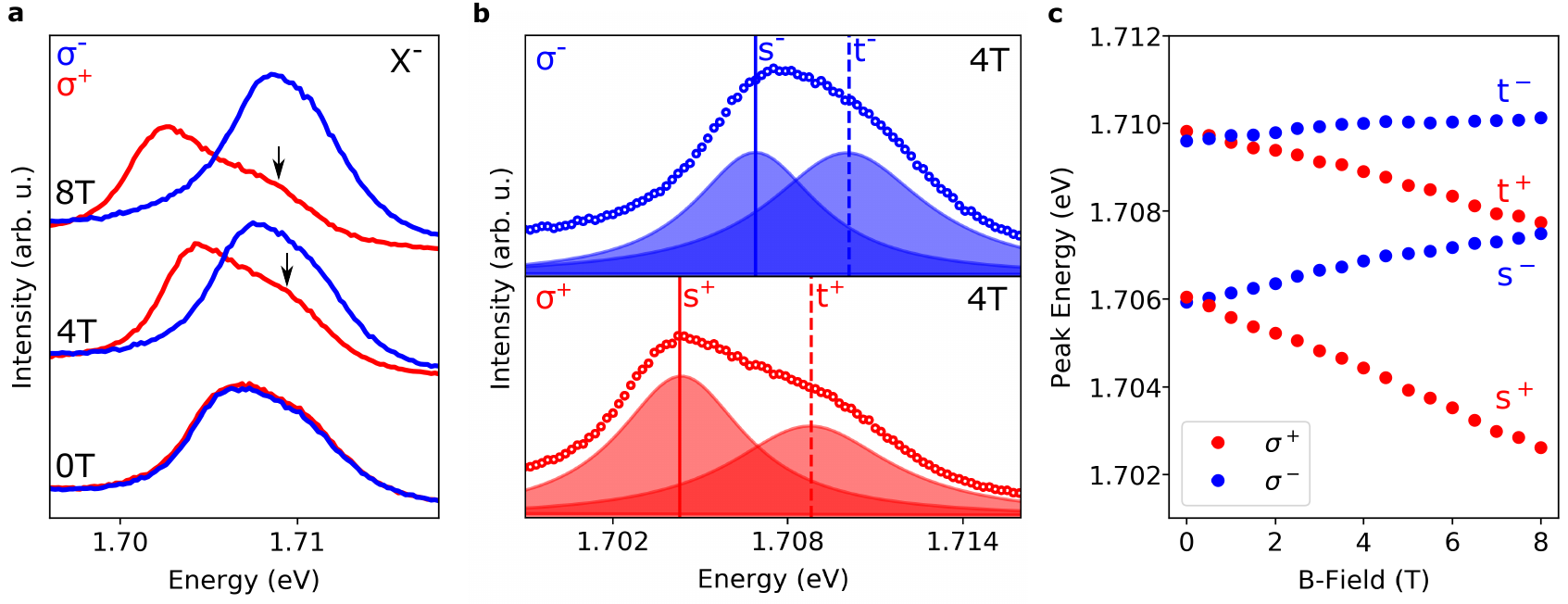}
	\caption{\label{fig:figure3} (a) Circular polarization resolved trion photoluminescence spectra at B = 0 T, 4 T, and 8 T. The energy separation between fine structure components appears dependent on B-field, as evidenced by the growing separation between the main peak and the shoulder (indicated by black arrows) in $\sigma^+$ polarization with increasing field strength. The increasing symmetry and intensity of the emission in $\sigma^-$ polarization suggests converging fine structure components with B. (b) The trion fine structure may be fitted to two Lorentzian peaks in each photon helicity, showing all four trion configurations at B = 4 T. Open circles are the CCD data. (c) Peak energy of the fitted fine structure components as a function of external B-field strength. The energy splitting between singlet trions is significantly larger than between triplet trions.}
\end{figure*}

At 30 K, and initially at zero external B-field, we observe in PL an asymmetric trion feature composed of an unresolved lower energy singlet peak and higher energy triplet peak, as can be seen in the 0 T trace of Fig.~\ref{fig:figure3}a. As depicted in Fig.~\ref{fig:figure1}, emission from $t^{+}$ ($s^{+}$) and $t^{-}$ ($s^{-}$) is at the same energy in the absence of an external B-field, and there is a few meV energy gap $\delta_{ex}$ between the triplet emission and singlet emission, arising from the intervalley exchange coupling.

Upon applying an external B-field perpendicular to the sample, up to B = 8 T, the $\sigma^{+}$ and $\sigma^{-}$ components of the emission shift spectrally away from one another, with $\sigma^{-}$ emission shifting to higher energy, consistent with the valley Zeeman effect. However, as is clear from the 4 T and 8 T traces of Fig.~\ref{fig:figure3}a, there is an accompanying lineshape evolution of the trion feature with external field. It appears that when shifting to lower energy, the singlet and triplet increase their energy separation, as evidenced by the prominent shoulder appearing in the $\sigma^{+}$ emission at B $>$ 0, highlighted by black arrows. Conversely, when shifting to higher energy, the singlet and triplet peaks appear to reduce their energy separation, resulting in the overall brighter and narrower emission profile seen in the $\sigma^{-}$ emission at B $>$ 0.

In order to extract the photon energies of the four different trion states, i.e. $s^{+}, s^{-}, t^{+}, t^{-},$ the trion fine structure at 30 K was fitted with two Lorentzian components for each PL polarization. An example of the fitted fine structure is shown in Fig.~\ref{fig:figure3}b, at B = 4 T, where it is clear that four distinct peaks exist in total, corresponding to the four ground state trion varieties. Each of these fitted Lorentzian peaks is distinct from the other three by either photon energy, helicity, or both. As such, it is possible to isolate and trace the energy shift of each trion state independently over the external B-field range, as shown in Fig.~\ref{fig:figure3}c. Here, several observations can be made. Firstly, subtraction of the $s^{+}$ peak energy from $t^{+}$, or equivalently $s^{-}$ from $t^{-}$, at B = 0 T, yields the energy difference $\delta_{ex}$. In this sample we find $\delta_{ex}\approx4$ meV, smaller than previously reported values of $\approx 6-7$ meV \cite{jones2016excitonic}. Secondly, there is a larger apparent valley splitting between singlet trions than between triplet trions, consistent with what may be inferred from the raw spectra in Fig.~\ref{fig:figure3}a. Thirdly, the $\sigma^-$ trions appear to have lower rates of shift than their $\sigma^+$ counterparts, in other words, $t^{-}$ ($s^{-}$) is less sensitive to the external field than $t^{+}$ ($s^{+}$). 

To understand this behaviour, it is useful to consider the initial and final states of trion radiative recombination, under the influence of an external magnetic field in the Faraday geometry. The initial state consists of either a singlet trion, or a triplet trion at a raised energy $\delta_{ex}$. The additional electron present in each of these trion complexes ensures the initial state charge $=-1$, which quantizes the trion dispersion into Landau levels (LLs), with a cyclotron frequency $\omega_{X^-} = e B / m_{X^-}$ where $e$ is the electron charge and $m_{X^-}$ the trion effective mass. Furthermore, the initial state is subject to energy shifts arising from the atomic orbital and Berry curvature associated magnetic moments inherent to monolayer WSe$_2$, much like the neutral exciton \cite{aivazian2015magnetic,srivastava2015valley,wang2015magneto, yu2014dirac}. The atomic orbitals constituting the valence band edge have a magnetic moment of magnitude $2 \mu_B$, which leads to an expected valley splitting of magnitude $4 \mu_B B$. Any discrepancy from this value arises due to the Berry curvature, or ``valley" magnetic moment, which acts on the valley psuedospin. In analogy to the neutral exciton valley Zeemen effect, we can express the energy shift of the initial state trion as $\frac{1}{2} \tau_z g_z \mu_B B$ where $\tau_z = \pm1$ for $\sigma^{\pm}$ emission helicity, and $g_z$ describes the cumulative effect of atomic orbital and Berry curvature magnetic moments.

From the initial state $t^{-}$ or $s^{+}$ ($t^{+}$ or $s^{-}$), the final state after trion recombination will be a photon and a single electron in the conduction band state $\ket{+K,\downarrow}$ ($\ket{-K,\uparrow}$). In each of these two final states, the electron experiences magnetic moments due to both the spin and valley pseudospin, which counteract one another. The cumulative ``spin-valley" electron g-factor $g_e$ in band $c1$ therefore depends on the relative strengths of these two opposing magnetic moments. The energy shift of an electron in band $c1$ may be expressed as $\frac{1}{2} \tau_e g_e \mu_B B$ where $\tau_e = \pm1$ for the electron in the $\pm$K valley, as a consequence of time reversal symmetry. 

In addition to the spin and valley energy shifts, the final state electron is also subject to LL quantization, however, the electron cyclotron energy $\omega_e$ will be much larger than $\omega_{X^-}$ thanks to the much smaller electron effective mass. Consequently, when a trion radiatively recombines, the additional energy of the electron LL relative to the trion LL is deducted from the photon energy. This leads to a global redshift of trion PL with increasing B-field strength, which may be quantified by an effective g-factor $g_l$ as $\hbar \omega_e - \hbar \omega_{X^-} = g_l \mu_B B$. In Fig.~\ref{fig:figure3}c the redshift manifests as the shallower gradient of $t^{-}$ ($s^{-}$) relative to $t^{+}$ ($s^{+}$).

\begin{figure}
	\center
	\includegraphics{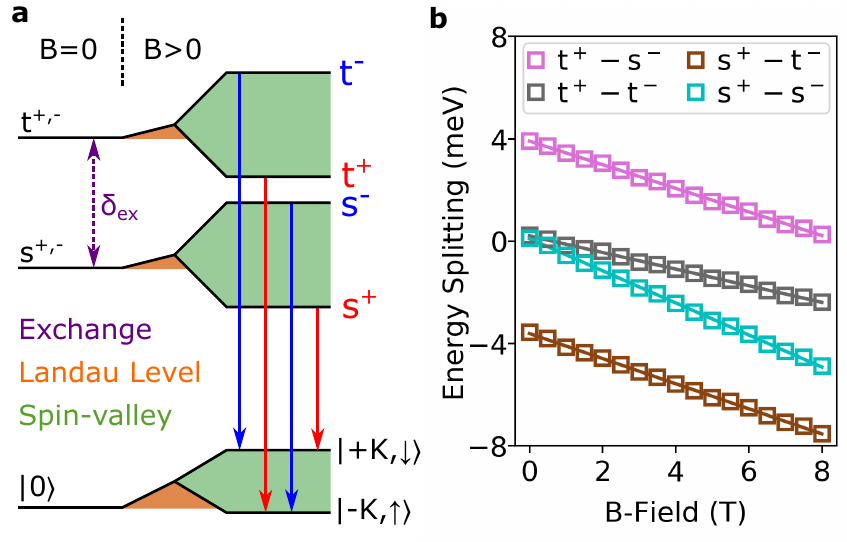}
	\caption{\label{fig:figure4} (a) Energy level diagram showing initial (one of the four trion varieties) and final (photon plus single electron in band $c1$) states of trion optical recombination. The intervalley electron-hole exchange interaction creates the energy gap $\delta_{ex}$, lifting the degeneracy between singlet and triplet trions. Under a positive external B-field, further energy shifts arise from Landau level quantization of both trions and free electrons (orange areas), along with spin and valley associated magnetic moments (green areas). See main text for details. Energies are not to scale. (b) Energy separation as a function of external B-field between oppositely circularly polarized trion configurations, calculated from $E(\sigma^+) - E(\sigma^-)$ in photoluminescence.}
\end{figure}

Overall, we define the change in emitted photon energy $\Delta E_{h\nu}$ as a function of the change in external magnetic field, $\Delta B$ as

\begin{equation}
\Delta E_{h\nu} = \frac{1}{2}(\tau_z g_z - \tau_e g_e - 2g_l) \mu_B \Delta B
\end{equation}

where $g_z$ may be viewed as the excitonic valley Zeeman g-factor of the trion, equal for all trion states, and $g_e$ and $g_l$ are modifications to the emitted photon energy arising purely from the recoil energy of the excess electron. Unlike neutral exciton recombination, trion recombination leaves behind an electron with non-zero momentum, which detracts energy from the emitted photon. Such recoil processes are therefore trion specific, and give rise to the asymmetric energy splittings shown in Fig.~\ref{fig:figure3}c. Each of the energy shifts in eq. (1) are shown schematically in Fig.~\ref{fig:figure4}a, along with the relative photon energies when B $>0$. We note that in order to reproduce the data in Fig.~\ref{fig:figure3}c, the band $c1$ valley magnetic moment must have larger magnitude than the spin magnetic moment, such that under a positive external B-field, the state $\ket{+K,\downarrow}$ is at higher energy than $\ket{-K,\uparrow}$. 

In order to extract these various g-factors from the magneto-PL measurements, we consider the photon energy separations between trions of opposite PL polarization, in the convention $E(\sigma^+) - E(\sigma^-)$, as plotted in Fig.~\ref{fig:figure4}b. Remarkably, despite the complexities of three distinct g-factors acting on four distinct trion states, the inherent symmetries in the system cause the energy splittings to become quite simplistic. Table~\ref{tab:table1} lists the measured gradients of each line in Fig.~\ref{fig:figure4}b, and the corresponding description calculated from eq. (1). By taking the mean value of Table~\ref{tab:table1} rows 1 and 2, we obtain $g_z = -8.3\pm0.2$, corresponding to a trion valley Zeeman splitting of $\approx -8.3 \mu_B B$. This is approximately double the value expected from purely atomic orbital contributions in the valence band ($-4 \mu_B B$), implying a large Berry curvature associated magnetic moment, in agreement with previous suggestions \cite{srivastava2015valley,plechinger2016excitonic,yu2014dirac}. The opening of the energy gap $\delta_{ex}$ between oppositely circularly polarized dispersion minima transforms the trion into a massive Dirac particle, associated with a large Berry curvature $\Omega(\textbf{k})$ \cite{yu2014dirac,xu2014spin}. The contribution to $g_z$ from the Berry curvature may be expressed as $\frac{m_e}{2\hbar^2} \delta_{ex} \Omega(\textbf{k})$ \cite{srivastava2015valley} (see Supplementary Note 1). Our data suggest that this contribution amounts to $\sim 4$ (as $|g_z| \approx 8$ and the atomic orbital contribution $\sim 4$), which yields a value of $\Omega(\pm K) \sim 10^4$ \AA$^2$. This is in excellent agreement with the predicted value when modelling the trion as a massive Dirac fermion \cite{yu2014dirac}.

\begin{table}
	\centering
	\begin{tabular}{| c | c | c | c |}
		\hline
		Row & Energy Separation & Measured Gradient & Corresponds to:  \\ \hline \hline
		1&$E(s^+) - E(t^-)$ & $(-8.5\pm0.2)\mu_B$ & $g_z\mu_B$  \\ \hline
		2&$E(t^+) - E(s^-)$ & $(-8.0\pm0.2)\mu_B$ & $g_z\mu_B$  \\ \hline
		3&$E(t^+) - E(t^-)$ & $(-5.6\pm0.2)\mu_B$ & $(g_z + g_e)\mu_B$  \\ \hline
		4&$E(s^+) - E(s^-)$ & $(-10.9\pm0.2)\mu_B$ & $(g_z - g_e)\mu_B$ \\ \hline
	\end{tabular}
	\caption{\label{tab:table1} List of measured gradients extracted from data shown in Fig.~\ref{fig:figure4}b, and their representations in terms of $g_z$ and $g_e$ from eq. (1). Taking the mean of the gradients in rows 1 and 2 yields $g_z = -8.3\pm0.2$, and using this value in conjuction with the gradients in rows 3 and 4 yields a mean $g_e = 2.7 \pm0.2$. See main text for details.}
\end{table}

From Table~\ref{tab:table1} we also extract $g_e$ by taking the mean value of rows 3 and 4 with $g_z = -8.3\pm0.2$. The result is a valley splitting of band $c1$ of $E_{\ket{+K,\downarrow}} - E_{\ket{-K,\uparrow}} = g_e \mu_B B = (2.7 \pm0.2) \mu_B B$, corresponding to a single electron magnetic moment of $\pm(1.4\pm0.1) \mu_B$ in the $\pm$K valley of band $c1$. Inserting $g_z = -8.3\pm0.2$ and $g_e = 2.7 \pm0.2$ into eq. (1) allows extraction of $g_l$ from the line gradients in Fig.~\ref{fig:figure3}c, as listed in Table~\ref{tab:table2}. Calculating the mean of the values from each trion line yields $g_l = 1.8\pm0.1$. Taking typical electron and trion effective masses from published density functional theory calculations yields a good agreement with the measured values of $g_l$ \cite{kormanyos2015k} (see Supplementary Note 1).

\begin{table}
	\centering
	\begin{tabular}{ | c | c | c |}
		\hline
		Trion State &$\; \; \; \;$ Measured Gradient $\; \; \; \;$ & $\; \; \; \;\; \; \; \; \; \; \; g_l \; \; \; \; \; \; \; \;\; \; \;$  \\ \hline \hline
		$t^-$ & $(1.2\pm0.2)\mu_B$  & $1.6\pm0.2$  \\ \hline
		$t^+$ & $(-4.6\pm0.1)\mu_B$  & $1.8\pm0.2$  \\ \hline
		$s^-$ & $(3.8\pm0.2)\mu_B$  & $1.7\pm0.2$  \\ \hline
		$s^+$ & $(-7.5\pm0.1)\mu_B$  & $2.0\pm0.2$ \\ \hline
	\end{tabular}
	\caption{\label{tab:table2} List of measured gradients extracted from data shown in Fig.~\ref{fig:figure3}c, and the corresponding value of $g_l$ calculated from eq. (1) when taking $g_z = -8.3\pm0.2$ and $g_e = 2.7 \pm0.2$. The mean value of $g_l$ is $1.8\pm0.1$. See main text for details.}
\end{table}

\section{Discussion}

In this work, we reveal the complexities of trion magneto-PL, demonstrating that the energy splitting in PL does not reflect the underlying valley Zeeman splitting of the initial state trion, as it does with the neutral exciton. Instead, the process of trion radiative recombination itself modifies the emitted photon energy, via the electron recoil, having the effect of enhancing the singlet-singlet splitting, and diminishing the triplet-triplet splitting. We arrive at the significant conclusion that any measurement of the trion valley splitting when treating it as a single resonance (without fine structure) cannot yield an accurate measurement of the true valley Zeeman effect, as the measured value will depend on the relative contributions of the four fine structure components, each of which have different rates of shift, as shown in Table~\ref{tab:table2}. As this work demonstrates, the relative spectral weighting of the singlets and triplets has a strong temperature dependence. To our knowledge, all as-yet published reports of the trion valley splitting in WSe$_2$ treat the trion as a single PL peak, and so it may be expected that the slightly different temperatures used in these studies will modify the overall measured g-factor in an unpredictable manner. Crucially, even at extremely low temperatures $\sim 4$ K, where the triplet emission could feasibly be neglected, the measurement is not simplified, as the singlet line shifts would still depend on all of $g_z$, $g_e$, and $g_l$.

To further complicate matters, the LL associated g-factor $g_l$ is likely to depend heavily on sample variation. In our model, we assume the trions and electrons occupy their respective $n=1$ LLs, in both $\pm$K valleys, which ensures that $g_l$ cancels out when measuring an energy separation between two trion states. In reality, however, the LL structure may not be so simple \cite{wang2017valley}. The random electron doping in any given monolayer may result in complete filling of the lowest order LLs, forcing the excess electron to occupy increasingly higher order LLs upon recoil. Under a magnetic field, the band $c1$ will evolve into a series of spin-valley contrasting states with overlapping energies, ultimately leading to different LL filling in opposite valleys, and causing $g_l$ to take valley-specific values. This will lead to additional modifications to the energy splitting between trion states of opposite excess electron valley index. The observed trion PL splitting will therefore depend on the exact nature of the interplay between the spin-valley polarized 2-dimensional electron gas, the LL filling, and temperature, which induces Fermi level broadening.

In conclusion, we exploit the unique valley symmetries of WSe$_2$ trions to optically measure all effective g-factors arising from valley Zeeman and electron recoil processes, which reveals that the valley Zeeman framework reported for neutral excitons is insufficient to describe the magneto-optical response of trions in WSe$_2$. From purely optical measurements, we extract the single electron magnetic moment in the band $c1$, which will be critical information for research regarding spin-valley currents in future valleytronic devices. Knowledge of the band $c1$ may also shed new light on the magnetic response of dark excitons in this material, the properties of which are highly elusive owing to their spin forbidden optical transition \cite{zhang2017magnetic}. The results presented here gain critical insight into the magneto-photoluminescence of trion fine structure in monolayer WSe$_2$, information which will be crucial in future research involving the spin and valley dynamics of monolayer TMDs and their applications in valleytronics. 
	
\section{Methods}

\subsection{Low temperature magneto-optical spectroscopy}
Low temperature magneto-photoluminescence spectroscopy was performed by mounting the sample in a liquid helium bath cryostat containing a sample heating element and superconducting magnet coil. Non-resonant continuous-wave excitation at 1.946 eV in either $\sigma^{+}$ or $\sigma^{-}$ polarization was used, along with helicity selective circularly polarized detection, guided to a spectrometer and high sensitivity CCD.

\subsection{Sample fabrication}	

The hBN/WSe$_2$/hBN stack was fabricated as follows. Firstly, bulk hBN crystals were mechanically exfoliated onto a polymer double layer commonly used for dry-transfer methods \cite{kretinin2014electronic}. The WSe$_2$ single-layer flake was then picked up from a separate Si/SiO$_2$  substrate using the hBN crystal on the poly(methyl methacrylate) (PMMA) membrane. This pick-up method was repeated to lift another thin hBN flake from a second Si/SiO$_2$ substrate.  The WSe$_2$ monolayer crystal is then fully protected from subsequent environmental degradation. The pick-up transfer was conducted with the target substrate held at T = 60$^\circ$C. The whole stack along with the PMMA membrane was then lowered onto a dielectric distributed Bragg reflector (DBR) substrate. The substrate consists of alternating layers of Ta$_2$O$_5$ and SiO$_2$ of $\sim$100 nm thickness, with SiO$_2$ as the top layer. The PMMA membrane along with the heterostructure stack was heated to 130$^\circ$C to soften the PMMA. Subsequent electron beam lithography and metallization was carried out to mechanically clamp as well as aid the locating of the heterostructure on the substrate.

\section{Acknowledgements}

The authors thank the financial support of the EPSRC grants EP/P026850/1, EP/M012727/1, EP/N031776/1 and EP/J007544/1, the European Union's Horizon 2020 research and innovation programme under ITN Spin-NANO Marie Sklodowska-Curie grant agreement 676108, and the Graphene Flagship under grant agreements 696656 and 785219. K. S. N. thanks financial support from the Royal Society, EPSRC, US Army Research Office (W911NF-16-1-0279) and ERC Grant Hetero2D. K.W. and T.T. acknowledge support from the Elemental Strategy Initiative conducted by the MEXT, Japan and and the CREST (JPMJCR15F3), JST.

\section{Author Contributions}

T. P. L. and S. D. carried out optical investigations. F. W. fabricated the sample. T. T. and K. W. synthesised hBN crystals. T. P. L., S. D. and A. I. T. analyzed the data. M. B., T. P. L. and G. B. developed a theoretical framework to interpret the data. K. S. N., G. B. and A. I. T. provided management of various aspects of the project. T. P. L. wrote the manuscript with contributions from all co-authors. T. P. L., S. D. and A. I. T. conceived the experiments. A. I. T. supervised the project.

\section{Competing Interests}

The authors declare no competing interests.
	

\begin{thebibliography}{31}%
	\makeatletter
	\providecommand \@ifxundefined [1]{%
		\@ifx{#1\undefined}
	}%
	\providecommand \@ifnum [1]{%
		\ifnum #1\expandafter \@firstoftwo
		\else \expandafter \@secondoftwo
		\fi
	}%
	\providecommand \@ifx [1]{%
		\ifx #1\expandafter \@firstoftwo
		\else \expandafter \@secondoftwo
		\fi
	}%
	\providecommand \natexlab [1]{#1}%
	\providecommand \enquote  [1]{``#1''}%
	\providecommand \bibnamefont  [1]{#1}%
	\providecommand \bibfnamefont [1]{#1}%
	\providecommand \citenamefont [1]{#1}%
	\providecommand \href@noop [0]{\@secondoftwo}%
	\providecommand \href [0]{\begingroup \@sanitize@url \@href}%
	\providecommand \@href[1]{\@@startlink{#1}\@@href}%
	\providecommand \@@href[1]{\endgroup#1\@@endlink}%
	\providecommand \@sanitize@url [0]{\catcode `\\12\catcode `\$12\catcode
		`\&12\catcode `\#12\catcode `\^12\catcode `\_12\catcode `\%12\relax}%
	\providecommand \@@startlink[1]{}%
	\providecommand \@@endlink[0]{}%
	\providecommand \url  [0]{\begingroup\@sanitize@url \@url }%
	\providecommand \@url [1]{\endgroup\@href {#1}{\urlprefix }}%
	\providecommand \urlprefix  [0]{URL }%
	\providecommand \Eprint [0]{\href }%
	\providecommand \doibase [0]{http://dx.doi.org/}%
	\providecommand \selectlanguage [0]{\@gobble}%
	\providecommand \bibinfo  [0]{\@secondoftwo}%
	\providecommand \bibfield  [0]{\@secondoftwo}%
	\providecommand \translation [1]{[#1]}%
	\providecommand \BibitemOpen [0]{}%
	\providecommand \bibitemStop [0]{}%
	\providecommand \bibitemNoStop [0]{.\EOS\space}%
	\providecommand \EOS [0]{\spacefactor3000\relax}%
	\providecommand \BibitemShut  [1]{\csname bibitem#1\endcsname}%
	\let\auto@bib@innerbib\@empty
	\bibitem [{\citenamefont {Wang}\ \emph {et~al.}(2018)\citenamefont {Wang},
		\citenamefont {Chernikov}, \citenamefont {Glazov}, \citenamefont {Heinz},
		\citenamefont {Marie}, \citenamefont {Amand},\ and\ \citenamefont
		{Urbaszek}}]{wang2018colloquium}%
	\BibitemOpen
	\bibfield  {author} {\bibinfo {author} {\bibfnamefont {G}~\bibnamefont
			{Wang}}, \bibinfo {author} {\bibfnamefont {A}~\bibnamefont {Chernikov}},
		\bibinfo {author} {\bibfnamefont {M~M}\ \bibnamefont {Glazov}}, \bibinfo
		{author} {\bibfnamefont {T~F}\ \bibnamefont {Heinz}}, \bibinfo {author}
		{\bibfnamefont {X}~\bibnamefont {Marie}}, \bibinfo {author} {\bibfnamefont
			{T}~\bibnamefont {Amand}}, \ and\ \bibinfo {author} {\bibfnamefont
			{B}~\bibnamefont {Urbaszek}},\ }\bibfield  {title} {\enquote {\bibinfo
			{title} {Colloquium: Excitons in atomically thin transition metal
				dichalcogenides},}\ }\href@noop {} {\bibfield  {journal} {\bibinfo  {journal}
			{Reviews of Modern Physics}\ }\textbf {\bibinfo {volume} {90}},\ \bibinfo
		{pages} {021001} (\bibinfo {year} {2018})}\BibitemShut {NoStop}%
	\bibitem [{\citenamefont {Mak}\ \emph {et~al.}(2010)\citenamefont {Mak},
		\citenamefont {Lee}, \citenamefont {Hone}, \citenamefont {Shan},\ and\
		\citenamefont {Heinz}}]{mak2010atomically}%
	\BibitemOpen
	\bibfield  {author} {\bibinfo {author} {\bibfnamefont {K~F}\ \bibnamefont
			{Mak}}, \bibinfo {author} {\bibfnamefont {C}~\bibnamefont {Lee}}, \bibinfo
		{author} {\bibfnamefont {J}~\bibnamefont {Hone}}, \bibinfo {author}
		{\bibfnamefont {J}~\bibnamefont {Shan}}, \ and\ \bibinfo {author}
		{\bibfnamefont {T~F}\ \bibnamefont {Heinz}},\ }\bibfield  {title} {\enquote
		{\bibinfo {title} {Atomically thin {MoS}$_2$: a new direct-gap
				semiconductor},}\ }\href@noop {} {\bibfield  {journal} {\bibinfo  {journal}
			{Physical Review Letters}\ }\textbf {\bibinfo {volume} {105}},\ \bibinfo
		{pages} {136805} (\bibinfo {year} {2010})}\BibitemShut {NoStop}%
	\bibitem [{\citenamefont {Splendiani}\ \emph {et~al.}(2010)\citenamefont
		{Splendiani}, \citenamefont {Sun}, \citenamefont {Zhang}, \citenamefont {Li},
		\citenamefont {Kim}, \citenamefont {Chim}, \citenamefont {Galli},\ and\
		\citenamefont {Wang}}]{splendiani2010emerging}%
	\BibitemOpen
	\bibfield  {author} {\bibinfo {author} {\bibfnamefont {A}~\bibnamefont
			{Splendiani}}, \bibinfo {author} {\bibfnamefont {L}~\bibnamefont {Sun}},
		\bibinfo {author} {\bibfnamefont {Y}~\bibnamefont {Zhang}}, \bibinfo {author}
		{\bibfnamefont {T}~\bibnamefont {Li}}, \bibinfo {author} {\bibfnamefont
			{J}~\bibnamefont {Kim}}, \bibinfo {author} {\bibfnamefont {C-Y}\ \bibnamefont
			{Chim}}, \bibinfo {author} {\bibfnamefont {G}~\bibnamefont {Galli}}, \ and\
		\bibinfo {author} {\bibfnamefont {F}~\bibnamefont {Wang}},\ }\bibfield
	{title} {\enquote {\bibinfo {title} {Emerging photoluminescence in monolayer
				{MoS}$_2$},}\ }\href@noop {} {\bibfield  {journal} {\bibinfo  {journal} {Nano
				letters}\ }\textbf {\bibinfo {volume} {10}},\ \bibinfo {pages} {1271--1275}
		(\bibinfo {year} {2010})}\BibitemShut {NoStop}%
	\bibitem [{\citenamefont {Yu}\ \emph {et~al.}(2014{\natexlab{a}})\citenamefont
		{Yu}, \citenamefont {Cui}, \citenamefont {Xu},\ and\ \citenamefont
		{Yao}}]{yu2014valley}%
	\BibitemOpen
	\bibfield  {author} {\bibinfo {author} {\bibfnamefont {H}~\bibnamefont {Yu}},
		\bibinfo {author} {\bibfnamefont {X}~\bibnamefont {Cui}}, \bibinfo {author}
		{\bibfnamefont {X}~\bibnamefont {Xu}}, \ and\ \bibinfo {author}
		{\bibfnamefont {W}~\bibnamefont {Yao}},\ }\bibfield  {title} {\enquote
		{\bibinfo {title} {Valley excitons in two-dimensional semiconductors},}\
	}\href@noop {} {\bibfield  {journal} {\bibinfo  {journal} {National Science
				Review}\ ,\ \bibinfo {pages} {nwu078}} (\bibinfo {year}
		{2014}{\natexlab{a}})}\BibitemShut {NoStop}%
	\bibitem [{\citenamefont {Xu}\ \emph {et~al.}(2014)\citenamefont {Xu},
		\citenamefont {Yao}, \citenamefont {Xiao},\ and\ \citenamefont
		{Heinz}}]{xu2014spin}%
	\BibitemOpen
	\bibfield  {author} {\bibinfo {author} {\bibfnamefont {X}~\bibnamefont {Xu}},
		\bibinfo {author} {\bibfnamefont {W}~\bibnamefont {Yao}}, \bibinfo {author}
		{\bibfnamefont {D}~\bibnamefont {Xiao}}, \ and\ \bibinfo {author}
		{\bibfnamefont {T~F}\ \bibnamefont {Heinz}},\ }\bibfield  {title} {\enquote
		{\bibinfo {title} {Spin and pseudospins in layered transition metal
				dichalcogenides},}\ }\href@noop {} {\bibfield  {journal} {\bibinfo  {journal}
			{Nature Physics}\ }\textbf {\bibinfo {volume} {10}},\ \bibinfo {pages}
		{343--350} (\bibinfo {year} {2014})}\BibitemShut {NoStop}%
	\bibitem [{\citenamefont {Mak}\ \emph {et~al.}(2012)\citenamefont {Mak},
		\citenamefont {He}, \citenamefont {Shan},\ and\ \citenamefont
		{Heinz}}]{mak2012control}%
	\BibitemOpen
	\bibfield  {author} {\bibinfo {author} {\bibfnamefont {K~F}\ \bibnamefont
			{Mak}}, \bibinfo {author} {\bibfnamefont {K}~\bibnamefont {He}}, \bibinfo
		{author} {\bibfnamefont {J}~\bibnamefont {Shan}}, \ and\ \bibinfo {author}
		{\bibfnamefont {T~F}\ \bibnamefont {Heinz}},\ }\bibfield  {title} {\enquote
		{\bibinfo {title} {Control of valley polarization in monolayer {MoS}$_2$ by
				optical helicity},}\ }\href@noop {} {\bibfield  {journal} {\bibinfo
			{journal} {Nature nanotechnology}\ }\textbf {\bibinfo {volume} {7}},\
		\bibinfo {pages} {494--498} (\bibinfo {year} {2012})}\BibitemShut {NoStop}%
	\bibitem [{\citenamefont {Jones}\ \emph {et~al.}(2013)\citenamefont {Jones},
		\citenamefont {Yu}, \citenamefont {Ghimire}, \citenamefont {Wu},
		\citenamefont {Aivazian}, \citenamefont {Ross}, \citenamefont {Zhao},
		\citenamefont {Yan}, \citenamefont {Mandrus}, \citenamefont {Xiao} \emph
		{et~al.}}]{jones2013optical}%
	\BibitemOpen
	\bibfield  {author} {\bibinfo {author} {\bibfnamefont {A~M}\ \bibnamefont
			{Jones}}, \bibinfo {author} {\bibfnamefont {H}~\bibnamefont {Yu}}, \bibinfo
		{author} {\bibfnamefont {N~J}\ \bibnamefont {Ghimire}}, \bibinfo {author}
		{\bibfnamefont {S}~\bibnamefont {Wu}}, \bibinfo {author} {\bibfnamefont
			{G}~\bibnamefont {Aivazian}}, \bibinfo {author} {\bibfnamefont {J~S}\
			\bibnamefont {Ross}}, \bibinfo {author} {\bibfnamefont {B}~\bibnamefont
			{Zhao}}, \bibinfo {author} {\bibfnamefont {J}~\bibnamefont {Yan}}, \bibinfo
		{author} {\bibfnamefont {D~G}\ \bibnamefont {Mandrus}}, \bibinfo {author}
		{\bibfnamefont {D}~\bibnamefont {Xiao}},  \emph {et~al.},\ }\bibfield
	{title} {\enquote {\bibinfo {title} {Optical generation of excitonic valley
				coherence in monolayer {WSe}$_2$},}\ }\href@noop {} {\bibfield  {journal}
		{\bibinfo  {journal} {Nature nanotechnology}\ }\textbf {\bibinfo {volume}
			{8}},\ \bibinfo {pages} {634--638} (\bibinfo {year} {2013})}\BibitemShut
	{NoStop}%
	\bibitem [{\citenamefont {Wang}\ \emph {et~al.}(2014)\citenamefont {Wang},
		\citenamefont {Bouet}, \citenamefont {Lagarde}, \citenamefont {Vidal},
		\citenamefont {Balocchi}, \citenamefont {Amand}, \citenamefont {Marie},\ and\
		\citenamefont {Urbaszek}}]{wang2014valley}%
	\BibitemOpen
	\bibfield  {author} {\bibinfo {author} {\bibfnamefont {G}~\bibnamefont
			{Wang}}, \bibinfo {author} {\bibfnamefont {L}~\bibnamefont {Bouet}}, \bibinfo
		{author} {\bibfnamefont {D}~\bibnamefont {Lagarde}}, \bibinfo {author}
		{\bibfnamefont {M}~\bibnamefont {Vidal}}, \bibinfo {author} {\bibfnamefont
			{A}~\bibnamefont {Balocchi}}, \bibinfo {author} {\bibfnamefont
			{T}~\bibnamefont {Amand}}, \bibinfo {author} {\bibfnamefont {X}~\bibnamefont
			{Marie}}, \ and\ \bibinfo {author} {\bibfnamefont {B}~\bibnamefont
			{Urbaszek}},\ }\bibfield  {title} {\enquote {\bibinfo {title} {Valley
				dynamics probed through charged and neutral exciton emission in monolayer
				{WSe}$_2$},}\ }\href@noop {} {\bibfield  {journal} {\bibinfo  {journal}
			{Physical Review B}\ }\textbf {\bibinfo {volume} {90}},\ \bibinfo {pages}
		{075413} (\bibinfo {year} {2014})}\BibitemShut {NoStop}%
	\bibitem [{\citenamefont {Mak}\ \emph {et~al.}(2013)\citenamefont {Mak},
		\citenamefont {He}, \citenamefont {Lee}, \citenamefont {Lee}, \citenamefont
		{Hone}, \citenamefont {Heinz},\ and\ \citenamefont {Shan}}]{mak2013tightly}%
	\BibitemOpen
	\bibfield  {author} {\bibinfo {author} {\bibfnamefont {K~F}\ \bibnamefont
			{Mak}}, \bibinfo {author} {\bibfnamefont {K}~\bibnamefont {He}}, \bibinfo
		{author} {\bibfnamefont {C}~\bibnamefont {Lee}}, \bibinfo {author}
		{\bibfnamefont {G~H}\ \bibnamefont {Lee}}, \bibinfo {author} {\bibfnamefont
			{J}~\bibnamefont {Hone}}, \bibinfo {author} {\bibfnamefont {T~F}\
			\bibnamefont {Heinz}}, \ and\ \bibinfo {author} {\bibfnamefont
			{J}~\bibnamefont {Shan}},\ }\bibfield  {title} {\enquote {\bibinfo {title}
			{Tightly bound trions in monolayer {MoS}$_2$},}\ }\href@noop {} {\bibfield
		{journal} {\bibinfo  {journal} {Nature materials}\ }\textbf {\bibinfo
			{volume} {12}},\ \bibinfo {pages} {207} (\bibinfo {year} {2013})}\BibitemShut
	{NoStop}%
	\bibitem [{\citenamefont {You}\ \emph {et~al.}(2015)\citenamefont {You},
		\citenamefont {Zhang}, \citenamefont {Berkelbach}, \citenamefont {Hybertsen},
		\citenamefont {Reichman},\ and\ \citenamefont {Heinz}}]{you2015observation}%
	\BibitemOpen
	\bibfield  {author} {\bibinfo {author} {\bibfnamefont {Y}~\bibnamefont
			{You}}, \bibinfo {author} {\bibfnamefont {X-X}\ \bibnamefont {Zhang}},
		\bibinfo {author} {\bibfnamefont {T~C}\ \bibnamefont {Berkelbach}}, \bibinfo
		{author} {\bibfnamefont {M~S}\ \bibnamefont {Hybertsen}}, \bibinfo {author}
		{\bibfnamefont {D~R}\ \bibnamefont {Reichman}}, \ and\ \bibinfo {author}
		{\bibfnamefont {T~F}\ \bibnamefont {Heinz}},\ }\bibfield  {title} {\enquote
		{\bibinfo {title} {Observation of biexcitons in monolayer {WS}e$_2$},}\
	}\href@noop {} {\bibfield  {journal} {\bibinfo  {journal} {Nature Physics}\
		}\textbf {\bibinfo {volume} {11}},\ \bibinfo {pages} {477} (\bibinfo {year}
		{2015})}\BibitemShut {NoStop}%
	\bibitem [{\citenamefont {Schaibley}\ \emph {et~al.}(2016)\citenamefont
		{Schaibley}, \citenamefont {Yu}, \citenamefont {Clark}, \citenamefont
		{Rivera}, \citenamefont {Ross}, \citenamefont {Seyler}, \citenamefont {Yao},\
		and\ \citenamefont {Xu}}]{schaibley2016valleytronics}%
	\BibitemOpen
	\bibfield  {author} {\bibinfo {author} {\bibfnamefont {J~R}\ \bibnamefont
			{Schaibley}}, \bibinfo {author} {\bibfnamefont {H}~\bibnamefont {Yu}},
		\bibinfo {author} {\bibfnamefont {G}~\bibnamefont {Clark}}, \bibinfo {author}
		{\bibfnamefont {P}~\bibnamefont {Rivera}}, \bibinfo {author} {\bibfnamefont
			{J~S}\ \bibnamefont {Ross}}, \bibinfo {author} {\bibfnamefont {K~L}\
			\bibnamefont {Seyler}}, \bibinfo {author} {\bibfnamefont {W}~\bibnamefont
			{Yao}}, \ and\ \bibinfo {author} {\bibfnamefont {X}~\bibnamefont {Xu}},\
	}\bibfield  {title} {\enquote {\bibinfo {title} {Valleytronics in 2d
				materials},}\ }\href@noop {} {\bibfield  {journal} {\bibinfo  {journal}
			{Nature Reviews Materials}\ }\textbf {\bibinfo {volume} {1}},\ \bibinfo
		{pages} {16055} (\bibinfo {year} {2016})}\BibitemShut {NoStop}%
	\bibitem [{\citenamefont {Xiao}\ \emph {et~al.}(2012)\citenamefont {Xiao},
		\citenamefont {Liu}, \citenamefont {Feng}, \citenamefont {Xu},\ and\
		\citenamefont {Yao}}]{xiao2012coupled}%
	\BibitemOpen
	\bibfield  {author} {\bibinfo {author} {\bibfnamefont {D}~\bibnamefont
			{Xiao}}, \bibinfo {author} {\bibfnamefont {G-B}\ \bibnamefont {Liu}},
		\bibinfo {author} {\bibfnamefont {W}~\bibnamefont {Feng}}, \bibinfo {author}
		{\bibfnamefont {X}~\bibnamefont {Xu}}, \ and\ \bibinfo {author}
		{\bibfnamefont {W}~\bibnamefont {Yao}},\ }\bibfield  {title} {\enquote
		{\bibinfo {title} {Coupled spin and valley physics in monolayers of {MoS}$_2$
				and other group-{VI} dichalcogenides},}\ }\href@noop {} {\bibfield  {journal}
		{\bibinfo  {journal} {Physical Review Letters}\ }\textbf {\bibinfo {volume}
			{108}},\ \bibinfo {pages} {196802} (\bibinfo {year} {2012})}\BibitemShut
	{NoStop}%
	\bibitem [{\citenamefont {Singh}\ \emph {et~al.}(2016)\citenamefont {Singh},
		\citenamefont {Tran}, \citenamefont {Kolarczik}, \citenamefont {Seifert},
		\citenamefont {Wang}, \citenamefont {Hao}, \citenamefont {Pleskot},
		\citenamefont {Gabor}, \citenamefont {Helmrich}, \citenamefont {Owschimikow}
		\emph {et~al.}}]{singh2016long}%
	\BibitemOpen
	\bibfield  {author} {\bibinfo {author} {\bibfnamefont {A}~\bibnamefont
			{Singh}}, \bibinfo {author} {\bibfnamefont {K}~\bibnamefont {Tran}}, \bibinfo
		{author} {\bibfnamefont {M}~\bibnamefont {Kolarczik}}, \bibinfo {author}
		{\bibfnamefont {J}~\bibnamefont {Seifert}}, \bibinfo {author} {\bibfnamefont
			{Y}~\bibnamefont {Wang}}, \bibinfo {author} {\bibfnamefont {K}~\bibnamefont
			{Hao}}, \bibinfo {author} {\bibfnamefont {D}~\bibnamefont {Pleskot}},
		\bibinfo {author} {\bibfnamefont {N~M}\ \bibnamefont {Gabor}}, \bibinfo
		{author} {\bibfnamefont {S}~\bibnamefont {Helmrich}}, \bibinfo {author}
		{\bibfnamefont {N}~\bibnamefont {Owschimikow}},  \emph {et~al.},\ }\bibfield
	{title} {\enquote {\bibinfo {title} {Long-lived valley polarization of
				intravalley trions in monolayer {WSe}$_2$},}\ }\href@noop {} {\bibfield
		{journal} {\bibinfo  {journal} {Physical Review Letters}\ }\textbf {\bibinfo
			{volume} {117}},\ \bibinfo {pages} {257402} (\bibinfo {year}
		{2016})}\BibitemShut {NoStop}%
	\bibitem [{\citenamefont {Dufferwiel}\ \emph {et~al.}(2017)\citenamefont
		{Dufferwiel}, \citenamefont {Lyons}, \citenamefont {Solnyshkov},
		\citenamefont {Trichet}, \citenamefont {Withers}, \citenamefont {Schwarz},
		\citenamefont {Malpuech}, \citenamefont {Smith}, \citenamefont {Novoselov},
		\citenamefont {Skolnick} \emph {et~al.}}]{dufferwiel2017valley}%
	\BibitemOpen
	\bibfield  {author} {\bibinfo {author} {\bibfnamefont {S}~\bibnamefont
			{Dufferwiel}}, \bibinfo {author} {\bibfnamefont {T~P}\ \bibnamefont {Lyons}},
		\bibinfo {author} {\bibfnamefont {D~D}\ \bibnamefont {Solnyshkov}}, \bibinfo
		{author} {\bibfnamefont {A~A~P}\ \bibnamefont {Trichet}}, \bibinfo {author}
		{\bibfnamefont {F}~\bibnamefont {Withers}}, \bibinfo {author} {\bibfnamefont
			{S}~\bibnamefont {Schwarz}}, \bibinfo {author} {\bibfnamefont
			{G}~\bibnamefont {Malpuech}}, \bibinfo {author} {\bibfnamefont {J~M}\
			\bibnamefont {Smith}}, \bibinfo {author} {\bibfnamefont {K~S}\ \bibnamefont
			{Novoselov}}, \bibinfo {author} {\bibfnamefont {M~S}\ \bibnamefont
			{Skolnick}},  \emph {et~al.},\ }\bibfield  {title} {\enquote {\bibinfo
			{title} {Valley-addressable polaritons in atomically thin semiconductors},}\
	}\href@noop {} {\bibfield  {journal} {\bibinfo  {journal} {Nature Photonics}\
		}\textbf {\bibinfo {volume} {11}},\ \bibinfo {pages} {497} (\bibinfo {year}
		{2017})}\BibitemShut {NoStop}%
	\bibitem [{\citenamefont {Aivazian}\ \emph {et~al.}(2015)\citenamefont
		{Aivazian}, \citenamefont {Gong}, \citenamefont {Jones}, \citenamefont {Chu},
		\citenamefont {Yan}, \citenamefont {Mandrus}, \citenamefont {Zhang},
		\citenamefont {Cobden}, \citenamefont {Yao},\ and\ \citenamefont
		{Xu}}]{aivazian2015magnetic}%
	\BibitemOpen
	\bibfield  {author} {\bibinfo {author} {\bibfnamefont {G}~\bibnamefont
			{Aivazian}}, \bibinfo {author} {\bibfnamefont {Z}~\bibnamefont {Gong}},
		\bibinfo {author} {\bibfnamefont {A~M}\ \bibnamefont {Jones}}, \bibinfo
		{author} {\bibfnamefont {R-L}\ \bibnamefont {Chu}}, \bibinfo {author}
		{\bibfnamefont {J}~\bibnamefont {Yan}}, \bibinfo {author} {\bibfnamefont
			{D~G}\ \bibnamefont {Mandrus}}, \bibinfo {author} {\bibfnamefont
			{C}~\bibnamefont {Zhang}}, \bibinfo {author} {\bibfnamefont {D}~\bibnamefont
			{Cobden}}, \bibinfo {author} {\bibfnamefont {W}~\bibnamefont {Yao}}, \ and\
		\bibinfo {author} {\bibfnamefont {X}~\bibnamefont {Xu}},\ }\bibfield  {title}
	{\enquote {\bibinfo {title} {Magnetic control of valley pseudospin in
				monolayer {WSe}$_2$},}\ }\href@noop {} {\bibfield  {journal} {\bibinfo
			{journal} {Nature Physics}\ } (\bibinfo {year} {2015})}\BibitemShut {NoStop}%
	\bibitem [{\citenamefont {Srivastava}\ \emph {et~al.}(2015)\citenamefont
		{Srivastava}, \citenamefont {Sidler}, \citenamefont {Allain}, \citenamefont
		{Lembke}, \citenamefont {Kis},\ and\ \citenamefont
		{Imamo{\u{g}}lu}}]{srivastava2015valley}%
	\BibitemOpen
	\bibfield  {author} {\bibinfo {author} {\bibfnamefont {A}~\bibnamefont
			{Srivastava}}, \bibinfo {author} {\bibfnamefont {M}~\bibnamefont {Sidler}},
		\bibinfo {author} {\bibfnamefont {A~V}\ \bibnamefont {Allain}}, \bibinfo
		{author} {\bibfnamefont {D~S}\ \bibnamefont {Lembke}}, \bibinfo {author}
		{\bibfnamefont {A}~\bibnamefont {Kis}}, \ and\ \bibinfo {author}
		{\bibfnamefont {A}~\bibnamefont {Imamo{\u{g}}lu}},\ }\bibfield  {title}
	{\enquote {\bibinfo {title} {Valley zeeman effect in elementary optical
				excitations of monolayer {WSe}$_2$},}\ }\href@noop {} {\bibfield  {journal}
		{\bibinfo  {journal} {Nature Physics}\ } (\bibinfo {year}
		{2015})}\BibitemShut {NoStop}%
	\bibitem [{\citenamefont {Wang}\ \emph {et~al.}(2015)\citenamefont {Wang},
		\citenamefont {Bouet}, \citenamefont {Glazov}, \citenamefont {Amand},
		\citenamefont {Ivchenko}, \citenamefont {Palleau}, \citenamefont {Marie},\
		and\ \citenamefont {Urbaszek}}]{wang2015magneto}%
	\BibitemOpen
	\bibfield  {author} {\bibinfo {author} {\bibfnamefont {G}~\bibnamefont
			{Wang}}, \bibinfo {author} {\bibfnamefont {L}~\bibnamefont {Bouet}}, \bibinfo
		{author} {\bibfnamefont {M~M}\ \bibnamefont {Glazov}}, \bibinfo {author}
		{\bibfnamefont {T}~\bibnamefont {Amand}}, \bibinfo {author} {\bibfnamefont
			{E~L}\ \bibnamefont {Ivchenko}}, \bibinfo {author} {\bibfnamefont
			{E}~\bibnamefont {Palleau}}, \bibinfo {author} {\bibfnamefont
			{X}~\bibnamefont {Marie}}, \ and\ \bibinfo {author} {\bibfnamefont
			{B}~\bibnamefont {Urbaszek}},\ }\bibfield  {title} {\enquote {\bibinfo
			{title} {Magneto-optics in transition metal diselenide monolayers},}\
	}\href@noop {} {\bibfield  {journal} {\bibinfo  {journal} {2D Materials}\
		}\textbf {\bibinfo {volume} {2}},\ \bibinfo {pages} {034002} (\bibinfo {year}
		{2015})}\BibitemShut {NoStop}%
	\bibitem [{\citenamefont {Koperski}\ \emph {et~al.}(2017)\citenamefont
		{Koperski}, \citenamefont {Molas}, \citenamefont {Arora}, \citenamefont
		{Nogajewski}, \citenamefont {Slobodeniuk}, \citenamefont {Faugeras},\ and\
		\citenamefont {Potemski}}]{koperski2017optical}%
	\BibitemOpen
	\bibfield  {author} {\bibinfo {author} {\bibfnamefont {M}~\bibnamefont
			{Koperski}}, \bibinfo {author} {\bibfnamefont {M~R}\ \bibnamefont {Molas}},
		\bibinfo {author} {\bibfnamefont {A}~\bibnamefont {Arora}}, \bibinfo {author}
		{\bibfnamefont {K}~\bibnamefont {Nogajewski}}, \bibinfo {author}
		{\bibfnamefont {A~O}\ \bibnamefont {Slobodeniuk}}, \bibinfo {author}
		{\bibfnamefont {C}~\bibnamefont {Faugeras}}, \ and\ \bibinfo {author}
		{\bibfnamefont {M}~\bibnamefont {Potemski}},\ }\bibfield  {title} {\enquote
		{\bibinfo {title} {Optical properties of atomically thin transition metal
				dichalcogenides: Observations and puzzles},}\ }\href@noop {} {\bibfield
		{journal} {\bibinfo  {journal} {Nanophotonics}\ } (\bibinfo {year}
		{2017})}\BibitemShut {NoStop}%
	\bibitem [{\citenamefont {Koperski}\ \emph {et~al.}(2015)\citenamefont
		{Koperski}, \citenamefont {Nogajewski}, \citenamefont {Arora}, \citenamefont
		{Cherkez}, \citenamefont {Mallet}, \citenamefont {Veuillen}, \citenamefont
		{Marcus}, \citenamefont {Kossacki},\ and\ \citenamefont
		{Potemski}}]{koperski2015single}%
	\BibitemOpen
	\bibfield  {author} {\bibinfo {author} {\bibfnamefont {M}~\bibnamefont
			{Koperski}}, \bibinfo {author} {\bibfnamefont {K}~\bibnamefont {Nogajewski}},
		\bibinfo {author} {\bibfnamefont {A}~\bibnamefont {Arora}}, \bibinfo {author}
		{\bibfnamefont {V}~\bibnamefont {Cherkez}}, \bibinfo {author} {\bibfnamefont
			{P}~\bibnamefont {Mallet}}, \bibinfo {author} {\bibfnamefont {J-Y}\
			\bibnamefont {Veuillen}}, \bibinfo {author} {\bibfnamefont {J}~\bibnamefont
			{Marcus}}, \bibinfo {author} {\bibfnamefont {P}~\bibnamefont {Kossacki}}, \
		and\ \bibinfo {author} {\bibfnamefont {M}~\bibnamefont {Potemski}},\
	}\bibfield  {title} {\enquote {\bibinfo {title} {Single photon emitters in
				exfoliated {WSe}$_2$ structures},}\ }\href@noop {} {\bibfield  {journal}
		{\bibinfo  {journal} {Nature nanotechnology}\ }\textbf {\bibinfo {volume}
			{10}},\ \bibinfo {pages} {503} (\bibinfo {year} {2015})}\BibitemShut
	{NoStop}%
	\bibitem [{\citenamefont {Yu}\ \emph {et~al.}(2014{\natexlab{b}})\citenamefont
		{Yu}, \citenamefont {Liu}, \citenamefont {Gong}, \citenamefont {Xu},\ and\
		\citenamefont {Yao}}]{yu2014dirac}%
	\BibitemOpen
	\bibfield  {author} {\bibinfo {author} {\bibfnamefont {H}~\bibnamefont {Yu}},
		\bibinfo {author} {\bibfnamefont {G-B}\ \bibnamefont {Liu}}, \bibinfo
		{author} {\bibfnamefont {P}~\bibnamefont {Gong}}, \bibinfo {author}
		{\bibfnamefont {X}~\bibnamefont {Xu}}, \ and\ \bibinfo {author}
		{\bibfnamefont {W}~\bibnamefont {Yao}},\ }\bibfield  {title} {\enquote
		{\bibinfo {title} {Dirac cones and dirac saddle points of bright excitons in
				monolayer transition metal dichalcogenides},}\ }\href@noop {} {\bibfield
		{journal} {\bibinfo  {journal} {Nature communications}\ }\textbf {\bibinfo
			{volume} {5}} (\bibinfo {year} {2014}{\natexlab{b}})}\BibitemShut {NoStop}%
	\bibitem [{\citenamefont {Jones}\ \emph {et~al.}(2016)\citenamefont {Jones},
		\citenamefont {Yu}, \citenamefont {Schaibley}, \citenamefont {Yan},
		\citenamefont {Mandrus}, \citenamefont {Taniguchi}, \citenamefont {Watanabe},
		\citenamefont {Dery}, \citenamefont {Yao},\ and\ \citenamefont
		{Xu}}]{jones2016excitonic}%
	\BibitemOpen
	\bibfield  {author} {\bibinfo {author} {\bibfnamefont {A~M}\ \bibnamefont
			{Jones}}, \bibinfo {author} {\bibfnamefont {H}~\bibnamefont {Yu}}, \bibinfo
		{author} {\bibfnamefont {J~R}\ \bibnamefont {Schaibley}}, \bibinfo {author}
		{\bibfnamefont {J}~\bibnamefont {Yan}}, \bibinfo {author} {\bibfnamefont
			{D~G}\ \bibnamefont {Mandrus}}, \bibinfo {author} {\bibfnamefont
			{T}~\bibnamefont {Taniguchi}}, \bibinfo {author} {\bibfnamefont
			{K}~\bibnamefont {Watanabe}}, \bibinfo {author} {\bibfnamefont
			{H}~\bibnamefont {Dery}}, \bibinfo {author} {\bibfnamefont {W}~\bibnamefont
			{Yao}}, \ and\ \bibinfo {author} {\bibfnamefont {X}~\bibnamefont {Xu}},\
	}\bibfield  {title} {\enquote {\bibinfo {title} {Excitonic luminescence
				upconversion in a two-dimensional semiconductor},}\ }\href@noop {} {\bibfield
		{journal} {\bibinfo  {journal} {Nature Physics}\ }\textbf {\bibinfo {volume}
			{12}},\ \bibinfo {pages} {323--327} (\bibinfo {year} {2016})}\BibitemShut
	{NoStop}%
	\bibitem [{\citenamefont {Plechinger}\ \emph
		{et~al.}(2016{\natexlab{a}})\citenamefont {Plechinger}, \citenamefont
		{Nagler}, \citenamefont {Arora}, \citenamefont {Schmidt}, \citenamefont
		{Chernikov}, \citenamefont {Grenados~del {\'A}guila}, \citenamefont
		{Christianen}, \citenamefont {Bratschitsch}, \citenamefont {Sch{\"u}ller},\
		and\ \citenamefont {Korn}}]{plechinger2016trion}%
	\BibitemOpen
	\bibfield  {author} {\bibinfo {author} {\bibfnamefont {G}~\bibnamefont
			{Plechinger}}, \bibinfo {author} {\bibfnamefont {P}~\bibnamefont {Nagler}},
		\bibinfo {author} {\bibfnamefont {A}~\bibnamefont {Arora}}, \bibinfo {author}
		{\bibfnamefont {R}~\bibnamefont {Schmidt}}, \bibinfo {author} {\bibfnamefont
			{A}~\bibnamefont {Chernikov}}, \bibinfo {author} {\bibfnamefont
			{A}~\bibnamefont {Grenados~del {\'A}guila}}, \bibinfo {author} {\bibfnamefont
			{P~C~M}\ \bibnamefont {Christianen}}, \bibinfo {author} {\bibfnamefont
			{R}~\bibnamefont {Bratschitsch}}, \bibinfo {author} {\bibfnamefont
			{C}~\bibnamefont {Sch{\"u}ller}}, \ and\ \bibinfo {author} {\bibfnamefont
			{T}~\bibnamefont {Korn}},\ }\bibfield  {title} {\enquote {\bibinfo {title}
			{Trion fine structure and coupled spin--valley dynamics in monolayer tungsten
				disulfide},}\ }\href@noop {} {\bibfield  {journal} {\bibinfo  {journal}
			{Nature Communications}\ }\textbf {\bibinfo {volume} {7}} (\bibinfo {year}
		{2016}{\natexlab{a}})}\BibitemShut {NoStop}%
	\bibitem [{\citenamefont {Vaclavkova}\ \emph {et~al.}(2018)\citenamefont
		{Vaclavkova}, \citenamefont {Wyzula}, \citenamefont {Nogajewski},
		\citenamefont {Bartos}, \citenamefont {Slobodeniuk}, \citenamefont
		{Faugeras}, \citenamefont {Potemski},\ and\ \citenamefont
		{Molas}}]{vaclavkova2018singlet}%
	\BibitemOpen
	\bibfield  {author} {\bibinfo {author} {\bibfnamefont {D}~\bibnamefont
			{Vaclavkova}}, \bibinfo {author} {\bibfnamefont {J}~\bibnamefont {Wyzula}},
		\bibinfo {author} {\bibfnamefont {K}~\bibnamefont {Nogajewski}}, \bibinfo
		{author} {\bibfnamefont {M}~\bibnamefont {Bartos}}, \bibinfo {author}
		{\bibfnamefont {A~O}\ \bibnamefont {Slobodeniuk}}, \bibinfo {author}
		{\bibfnamefont {C}~\bibnamefont {Faugeras}}, \bibinfo {author} {\bibfnamefont
			{M}~\bibnamefont {Potemski}}, \ and\ \bibinfo {author} {\bibfnamefont {M~R}\
			\bibnamefont {Molas}},\ }\bibfield  {title} {\enquote {\bibinfo {title}
			{Singlet and triplet trions in {WS}$_2$ monolayer encapsulated in hexagonal
				boron nitride},}\ }\href@noop {} {\bibfield  {journal} {\bibinfo  {journal}
			{Nanotechnology}\ }\textbf {\bibinfo {volume} {29}},\ \bibinfo {pages}
		{325705} (\bibinfo {year} {2018})}\BibitemShut {NoStop}%
	\bibitem [{\citenamefont {Plechinger}\ \emph
		{et~al.}(2016{\natexlab{b}})\citenamefont {Plechinger}, \citenamefont
		{Nagler}, \citenamefont {Arora}, \citenamefont {Grenados~del Águila},
		\citenamefont {Ballottin}, \citenamefont {Frank}, \citenamefont
		{Steinleitner}, \citenamefont {Gmitra}, \citenamefont {Fabian}, \citenamefont
		{Christianen} \emph {et~al.}}]{plechinger2016excitonic}%
	\BibitemOpen
	\bibfield  {author} {\bibinfo {author} {\bibfnamefont {G}~\bibnamefont
			{Plechinger}}, \bibinfo {author} {\bibfnamefont {P}~\bibnamefont {Nagler}},
		\bibinfo {author} {\bibfnamefont {A}~\bibnamefont {Arora}}, \bibinfo {author}
		{\bibfnamefont {A}~\bibnamefont {Grenados~del Águila}}, \bibinfo {author}
		{\bibfnamefont {M~V}\ \bibnamefont {Ballottin}}, \bibinfo {author}
		{\bibfnamefont {T}~\bibnamefont {Frank}}, \bibinfo {author} {\bibfnamefont
			{P}~\bibnamefont {Steinleitner}}, \bibinfo {author} {\bibfnamefont
			{M}~\bibnamefont {Gmitra}}, \bibinfo {author} {\bibfnamefont {J}~\bibnamefont
			{Fabian}}, \bibinfo {author} {\bibfnamefont {P~C~M}\ \bibnamefont
			{Christianen}},  \emph {et~al.},\ }\bibfield  {title} {\enquote {\bibinfo
			{title} {Excitonic valley effects in monolayer {WS}$_2$ under high magnetic
				fields},}\ }\href@noop {} {\bibfield  {journal} {\bibinfo  {journal} {Nano
				Letters}\ }\textbf {\bibinfo {volume} {16}},\ \bibinfo {pages} {7899--7904}
		(\bibinfo {year} {2016}{\natexlab{b}})}\BibitemShut {NoStop}%
	\bibitem [{\citenamefont {Taniguchi}\ and\ \citenamefont
		{Watanabe}(2007)}]{taniguchi2007synthesis}%
	\BibitemOpen
	\bibfield  {author} {\bibinfo {author} {\bibfnamefont {T}~\bibnamefont
			{Taniguchi}}\ and\ \bibinfo {author} {\bibfnamefont {K}~\bibnamefont
			{Watanabe}},\ }\bibfield  {title} {\enquote {\bibinfo {title} {Synthesis of
				high-purity boron nitride single crystals under high pressure by using
				{Ba--BN} solvent},}\ }\href@noop {} {\bibfield  {journal} {\bibinfo
			{journal} {Journal of crystal growth}\ }\textbf {\bibinfo {volume} {303}},\
		\bibinfo {pages} {525--529} (\bibinfo {year} {2007})}\BibitemShut {NoStop}%
	\bibitem [{\citenamefont {Cadiz}\ \emph {et~al.}(2017)\citenamefont {Cadiz},
		\citenamefont {Courtade}, \citenamefont {Robert}, \citenamefont {Wang},
		\citenamefont {Shen}, \citenamefont {Cai}, \citenamefont {Taniguchi},
		\citenamefont {Watanabe}, \citenamefont {Carrere}, \citenamefont {Lagarde}
		\emph {et~al.}}]{cadiz2017excitonic}%
	\BibitemOpen
	\bibfield  {author} {\bibinfo {author} {\bibfnamefont {F}~\bibnamefont
			{Cadiz}}, \bibinfo {author} {\bibfnamefont {E}~\bibnamefont {Courtade}},
		\bibinfo {author} {\bibfnamefont {C}~\bibnamefont {Robert}}, \bibinfo
		{author} {\bibfnamefont {G}~\bibnamefont {Wang}}, \bibinfo {author}
		{\bibfnamefont {Y}~\bibnamefont {Shen}}, \bibinfo {author} {\bibfnamefont
			{H}~\bibnamefont {Cai}}, \bibinfo {author} {\bibfnamefont {T}~\bibnamefont
			{Taniguchi}}, \bibinfo {author} {\bibfnamefont {K}~\bibnamefont {Watanabe}},
		\bibinfo {author} {\bibfnamefont {H}~\bibnamefont {Carrere}}, \bibinfo
		{author} {\bibfnamefont {D}~\bibnamefont {Lagarde}},  \emph {et~al.},\
	}\bibfield  {title} {\enquote {\bibinfo {title} {Excitonic linewidth
				approaching the homogeneous limit in {MoS}$_2$ based van der {W}aals
				heterostructures},}\ }\href@noop {} {\bibfield  {journal} {\bibinfo
			{journal} {Physical Review X}\ }\textbf {\bibinfo {volume} {7}},\ \bibinfo
		{pages} {021026} (\bibinfo {year} {2017})}\BibitemShut {NoStop}%
	\bibitem [{\citenamefont {Singh}\ \emph {et~al.}(2015)\citenamefont {Singh},
		\citenamefont {Knorr}, \citenamefont {Dass}, \citenamefont {Chen},
		\citenamefont {Malic}, \citenamefont {Moody}, \citenamefont {Clark},
		\citenamefont {Bergh{\"a}user}, \citenamefont {Hao}, \citenamefont {Tran}
		\emph {et~al.}}]{singh2015intrinsic}%
	\BibitemOpen
	\bibfield  {author} {\bibinfo {author} {\bibfnamefont {A}~\bibnamefont
			{Singh}}, \bibinfo {author} {\bibfnamefont {A}~\bibnamefont {Knorr}},
		\bibinfo {author} {\bibfnamefont {C~K}\ \bibnamefont {Dass}}, \bibinfo
		{author} {\bibfnamefont {C-H}\ \bibnamefont {Chen}}, \bibinfo {author}
		{\bibfnamefont {E}~\bibnamefont {Malic}}, \bibinfo {author} {\bibfnamefont
			{G}~\bibnamefont {Moody}}, \bibinfo {author} {\bibfnamefont {G}~\bibnamefont
			{Clark}}, \bibinfo {author} {\bibfnamefont {G}~\bibnamefont
			{Bergh{\"a}user}}, \bibinfo {author} {\bibfnamefont {K}~\bibnamefont {Hao}},
		\bibinfo {author} {\bibfnamefont {K}~\bibnamefont {Tran}},  \emph {et~al.},\
	}\bibfield  {title} {\enquote {\bibinfo {title} {Intrinsic homogeneous
				linewidth and broadening mechanisms of excitons in monolayer transition metal
				dichalcogenides},}\ }\href@noop {} {\bibfield  {journal} {\bibinfo  {journal}
			{Nature communications}\ }\textbf {\bibinfo {volume} {6}},\ \bibinfo {pages}
		{8315} (\bibinfo {year} {2015})}\BibitemShut {NoStop}%
	\bibitem [{\citenamefont {Korm{\'a}nyos}\ \emph {et~al.}(2015)\citenamefont
		{Korm{\'a}nyos}, \citenamefont {Burkard}, \citenamefont {Gmitra},
		\citenamefont {Fabian}, \citenamefont {Z{\'o}lyomi}, \citenamefont
		{Drummond},\ and\ \citenamefont {Fal’ko}}]{kormanyos2015k}%
	\BibitemOpen
	\bibfield  {author} {\bibinfo {author} {\bibfnamefont {A}~\bibnamefont
			{Korm{\'a}nyos}}, \bibinfo {author} {\bibfnamefont {G}~\bibnamefont
			{Burkard}}, \bibinfo {author} {\bibfnamefont {M}~\bibnamefont {Gmitra}},
		\bibinfo {author} {\bibfnamefont {J}~\bibnamefont {Fabian}}, \bibinfo
		{author} {\bibfnamefont {V}~\bibnamefont {Z{\'o}lyomi}}, \bibinfo {author}
		{\bibfnamefont {N~D}\ \bibnamefont {Drummond}}, \ and\ \bibinfo {author}
		{\bibfnamefont {V}~\bibnamefont {Fal’ko}},\ }\bibfield  {title} {\enquote
		{\bibinfo {title} {k{\textperiodcentered} p theory for two-dimensional
				transition metal dichalcogenide semiconductors},}\ }\href@noop {} {\bibfield
		{journal} {\bibinfo  {journal} {2D Materials}\ }\textbf {\bibinfo {volume}
			{2}},\ \bibinfo {pages} {022001} (\bibinfo {year} {2015})}\BibitemShut
	{NoStop}%
	\bibitem [{\citenamefont {Wang}\ \emph {et~al.}(2017)\citenamefont {Wang},
		\citenamefont {Shan},\ and\ \citenamefont {Mak}}]{wang2017valley}%
	\BibitemOpen
	\bibfield  {author} {\bibinfo {author} {\bibfnamefont {Z}~\bibnamefont
			{Wang}}, \bibinfo {author} {\bibfnamefont {J}~\bibnamefont {Shan}}, \ and\
		\bibinfo {author} {\bibfnamefont {K~F}\ \bibnamefont {Mak}},\ }\bibfield
	{title} {\enquote {\bibinfo {title} {Valley-and spin-polarized landau levels
				in monolayer {WSe}$_2$},}\ }\href@noop {} {\bibfield  {journal} {\bibinfo
			{journal} {Nature nanotechnology}\ }\textbf {\bibinfo {volume} {12}},\
		\bibinfo {pages} {144} (\bibinfo {year} {2017})}\BibitemShut {NoStop}%
	\bibitem [{\citenamefont {Zhang}\ \emph {et~al.}(2017)\citenamefont {Zhang},
		\citenamefont {Cao}, \citenamefont {Lu}, \citenamefont {Lin}, \citenamefont
		{Zhang}, \citenamefont {Wang}, \citenamefont {Li}, \citenamefont {Hone},
		\citenamefont {Robinson}, \citenamefont {Smirnov} \emph
		{et~al.}}]{zhang2017magnetic}%
	\BibitemOpen
	\bibfield  {author} {\bibinfo {author} {\bibfnamefont {X-X}\ \bibnamefont
			{Zhang}}, \bibinfo {author} {\bibfnamefont {T}~\bibnamefont {Cao}}, \bibinfo
		{author} {\bibfnamefont {Z}~\bibnamefont {Lu}}, \bibinfo {author}
		{\bibfnamefont {Y-C}\ \bibnamefont {Lin}}, \bibinfo {author} {\bibfnamefont
			{F}~\bibnamefont {Zhang}}, \bibinfo {author} {\bibfnamefont {Y}~\bibnamefont
			{Wang}}, \bibinfo {author} {\bibfnamefont {Z}~\bibnamefont {Li}}, \bibinfo
		{author} {\bibfnamefont {J~C}\ \bibnamefont {Hone}}, \bibinfo {author}
		{\bibfnamefont {J~A}\ \bibnamefont {Robinson}}, \bibinfo {author}
		{\bibfnamefont {D}~\bibnamefont {Smirnov}},  \emph {et~al.},\ }\bibfield
	{title} {\enquote {\bibinfo {title} {Magnetic brightening and control of dark
				excitons in monolayer {WSe}$_2$},}\ }\href@noop {} {\bibfield  {journal}
		{\bibinfo  {journal} {Nature nanotechnology}\ }\textbf {\bibinfo {volume}
			{12}},\ \bibinfo {pages} {883} (\bibinfo {year} {2017})}\BibitemShut
	{NoStop}%
	\bibitem [{\citenamefont {Kretinin}\ \emph {et~al.}(2014)\citenamefont
		{Kretinin}, \citenamefont {Cao}, \citenamefont {Tu}, \citenamefont {Yu},
		\citenamefont {Jalil}, \citenamefont {Novoselov}, \citenamefont {Haigh},
		\citenamefont {Gholinia}, \citenamefont {Mishchenko}, \citenamefont {Lozada}
		\emph {et~al.}}]{kretinin2014electronic}%
	\BibitemOpen
	\bibfield  {author} {\bibinfo {author} {\bibfnamefont {A~V}\ \bibnamefont
			{Kretinin}}, \bibinfo {author} {\bibfnamefont {Y}~\bibnamefont {Cao}},
		\bibinfo {author} {\bibfnamefont {J~S}\ \bibnamefont {Tu}}, \bibinfo {author}
		{\bibfnamefont {G~L}\ \bibnamefont {Yu}}, \bibinfo {author} {\bibfnamefont
			{R}~\bibnamefont {Jalil}}, \bibinfo {author} {\bibfnamefont {K~S}\
			\bibnamefont {Novoselov}}, \bibinfo {author} {\bibfnamefont {S~J}\
			\bibnamefont {Haigh}}, \bibinfo {author} {\bibfnamefont {A}~\bibnamefont
			{Gholinia}}, \bibinfo {author} {\bibfnamefont {A}~\bibnamefont {Mishchenko}},
		\bibinfo {author} {\bibfnamefont {M}~\bibnamefont {Lozada}},  \emph
		{et~al.},\ }\bibfield  {title} {\enquote {\bibinfo {title} {Electronic
				properties of graphene encapsulated with different two-dimensional atomic
				crystals},}\ }\href@noop {} {\bibfield  {journal} {\bibinfo  {journal} {Nano
				letters}\ }\textbf {\bibinfo {volume} {14}},\ \bibinfo {pages} {3270--3276}
		(\bibinfo {year} {2014})}\BibitemShut {NoStop}%
\end{thebibliography}
\end{document}


\title{Supplementary Information: The valley Zeeman effect in inter- and intra-valley trions in monolayer WSe$_2$}

\author{T. P. Lyons}
\email{tplyons1@sheffield.ac.uk}
\author{S. Dufferwiel}
\affiliation{Department of Physics and Astronomy, University of Sheffield, Sheffield S3 7RH, UK}
\author{M. Brooks}
\affiliation{Department of Physics, University of Konstanz, D-78464, Konstanz, Germany}
\author{F. Withers}
\affiliation{School of Physics and Astronomy, University of Manchester, Manchester M13 9PL, UK}
\affiliation{Centre for Graphene Science, CEMPS, University of Exeter, Exeter, EX4 4QF, UK}
\author{T. Taniguchi}
\affiliation{National Institute for Materials Science, Tsukuba, Ibaraki 305-0044, Japan}
\author{K. Watanabe}
\affiliation{National Institute for Materials Science, Tsukuba, Ibaraki 305-0044, Japan}
\author{K. S. Novoselov}
\affiliation{School of Physics and Astronomy, University of Manchester, Manchester M13 9PL, UK}
\author{G. Burkard}
\affiliation{Department of Physics, University of Konstanz, D-78464, Konstanz, Germany}
\author{A. I. Tartakovskii}
\email{a.tartakovskii@sheffield.ac.uk}
\affiliation{Department of Physics and Astronomy, University of Sheffield, Sheffield S3 7RH, UK}

\date{\today}

\maketitle

\subsection{Supplementary Note 1: A model for the trion valley Zeeman effect in monolayer WSe$_2$}

A negatively charged trion in a WSe$_2$ monolayer with a perpendicular magnetic field may be described by the following effective Hamiltonian

\begin{equation}
	\mathcal{H}=H_{T}+H_{\textrm{Int}}+H_{SO}+H_{Z}
\end{equation}

\noindent where $H_{T}$ describes the kinetic energy of all three constituent particles of the trion in a magnetic field, $H_{\textrm{Int}}$ describes the coulomb interactions, $H_{SO}$ the spin orbit splitting and $H_{Z}$ describes the Zeeman energies as follows

\begin{subequations}
	\begin{align}
		H_{T}&=\sum_{i=1}^{N=3}\left(-\frac{\hbar^2}{2 m_i}\boldsymbol{\nabla}_i+\frac{ e}{\hbar}\mathbf{A}_i\right)\\
		H_{\textrm{Int}}&=\sum_{i<j}^{N=3}V_{ij}(|\mathbf{r_i}-\mathbf{r_j}|)\\
		H_{SO}&=\tau_1 s_1 \Delta_{vb} + \sum_{i=2}^{N=3}\tau_i s_i \Delta_{cb}\\
		H_{Z}&=\frac{1}{2}(\tau_1 g_{vl}^{vb}+s_1 g_{s}^{vb}+\sum_{i=2}^{N=3}[\tau_i g_{vl}^{cb}+s_i g_{s}^{cb}])\mu_B B_Z.
	\end{align}
\end{subequations}

\noindent Here $i=1$ denotes the hole of the trion while $i=2,3$ denotes the bound and excess electron respectively. The effective masses and vector potential in the perpendicular magnetic field of the $i^{\textrm{th}}$ constituent particle is given by $m_i$ and $\mathbf{A}_i$ respectively. $V_{ij}(r)$ describes the interaction potential between charged particles, which in a TMD monolayer is best described with the Keldysh potential\cite{PhysRevB.96.085302} given as follows

\begin{equation}
	V_{ij}(r)=\frac{e_1e_2\pi}{2 r_0}\left[H_0\left(\frac{\kappa r}{r_0}\right)-Y_0\left(\frac{\kappa r}{r_0}\right)\right].
\end{equation}

\noindent Here $r_0$ is the screening length of the TMD, $\kappa$ is the average dielectric constant of the surrounding environment (in this case just the dielectric constant of hBN), $H_0$ is the $0^{\textrm{th}}$ Struve function and $Y_0$ is the $0^{\textrm{th}}$ Bessel Y function. Lastly, $g_{vl}^{vb/cb}$ is the valley g-factor in the valence/conduction band and $g_{s}^{vb/cb}$ is the spin g-factor in the valence/conduction band while $\tau_i=\pm1$ and $s_i=\pm1$ give the associated quantum number of the spin and valley pseudospin of the $i^{\textrm{th}}$ particle ($1$ corresponding to the $K$ valley or $\uparrow$ spin and $-1$ the $K'$ valley or $\downarrow$ spin).

The full expression for the change in emitted photon energy due to negative trion recombination in a magnetic field is 

\begin{equation}
	\Delta E_{h \nu}=\Delta E_{Z}-\Delta E_{\textrm{R}}+\Delta E_{\textrm{D}}
\end{equation}

\noindent where $\Delta E_{Z}$ is the Zeeman shift of the electron-hole pair within the trion, $\Delta E_{\textrm{R}}$ is the excess electron recoil shift and $\Delta E_{\textrm{D}}$ is the diamagnetic shift of the trion. The diamagnetic shift would induce a nonlinear shift in the observed trion lines with magnetic field. However, it is known from previous works\cite{Diamag65T,plechinger2016excitonic} that this nonlinearity is not expected to become relevant until field strengths of $>\unit[15]{T}$ are reached. As such, in this work, $\Delta E_{\textrm{D}}$ is not relevant.

Unlike exciton recombination, trion recombination cannot be a zero momentum process, as some momentum will be imparted to the excess electron as recoil, detracting from the observed photon energy. The excess electron recoil is dependent on both the temperature of the sample\cite{christopher2017long} and the Fermi level\cite{zhang2017recoil}, both of which are kept constant within this experiment. The effect of a perpendicular magnetic field on the excess electron recoil cost may be described by the difference in the trion and free electron first Landau level. Therefore the change in energy of the excess electron after trion recombination in a magnetic field is

\begin{equation}
	\Delta E^{\tau}_{e}=\frac{1}{2}\hbar \omega_{c}\left(1-\frac{m_3}{m_T}\right)
	+\frac{1}{2}\tau_{e} g_e\mu_B B_Z
\end{equation} 

\noindent where $\omega_{c}=e B_Z/m_3$ is the cyclotron frequency, $\tau_{e}$ is the valley index of the excess electron, $m_T=\sum_{i=1}^{N=3} m_i$ is the effective mass of the trion and $g_e$ is the bare out of plane g-factor of an electron in WSe$_2$. As the recoil energy cost is equal in the two valleys and this entire expression is linear in magnetic field, an effective valley dependent g-factor model can be used

\begin{equation}
	\Delta E^{\tau}_{e}=\frac{1}{2}\tau_{e} g_{\tau} \mu_B B_Z.
\end{equation}

\noindent Here $g_{\tau}$ is the valley dependent g-factors of the excess electron given in the model as

\begin{equation}
		g_{\tau}=g_e+\tau_{e} g_l
\end{equation}

\noindent where (5) and (6) may be solved to give a Landau level associated g-factor of 

\begin{equation}
	g_l=\frac{2 m_e m_X}{m_3  m_T}
\end{equation}

\noindent where $m_e$ is the bare electron mass and $m_X=\sum_{i=1}^{N=2} m_i$ is the exciton effective mass. Note the exciton mass here comes from the simplification of the Landau level component of (5) as we have assumed that the quasiparticle effective masses are linear combinations of the effective masses of the constituent particles. This accounts for the additional asymmetries of the observed photoluminescence energy gradients observed and detailed in Fig. 3c and Table II, but cancels out in the calculations of energy separations between different trion states, as shown in Table I of the main text. Using reasonable values of the effective masses for WSe$_2$ from Density Functional Theory (DFT) ($m_1=-0.36m_e$ and $m_2=m_3=0.29m_e$\cite{kormanyos2015k}) a $g_l=2.2$ close to the measured value ($\sim1.8$) is found.

Considering the spin-valley combinations of the four bright trions measured in this work, two triplet states $t^\pm$ and two singlet states $s^\pm$ addressed by $\sigma^\pm$ circularly polarised light, the Zeeman splitting may be reformulated as    

\begin{equation}
	H_{Z}=g_{t/s}^{\pm}\mu_B B_Z=\frac{1}{2}(\tau_z g_z-\tau_{e}g_e-2g_l)\mu_B B_Z.
\end{equation}

\noindent Here, $g_{t/s}^{\pm}$ is the g-factor of a specific variant of the bright trion measured, which in turn is composed of the exciton g-factor $g_z$ (given by the particles that will radiatively recombine) and the excess electron g-factor $g_e$, both with associated valley indices $\tau_z$ and $\tau_{e}$ respectively. As such, the total valley g-factors of each of the measured trions are as follows 

\begin{subequations}
	\begin{align}
		g_{s}^{+}&=\frac{1}{2}(g_z-g_e-2g_l)\\
		g_{s}^{-}&=-\frac{1}{2}(g_z-g_e+2g_l)\\
		g_{t}^{+}&=\frac{1}{2}(g_z+g_e-2g_l)\\
		g_{t}^{-}&=-\frac{1}{2}(g_z+g_e+2g_l).
	\end{align}
\end{subequations}

\noindent From this, the results shown in Table I. in the main text can be simply recovered 

\begin{subequations}
	\begin{align}
		g_{s}^{+}-g_{s}^{-}&=g_z-g_e\\
		g_{t}^{+}-g_{t}^{-}&=g_z+g_e\\
		g_{t}^{+}-g_{s}^{-}&=g_z\\
		g_{s}^{+}-g_{t}^{-}&=g_z.
	\end{align}
\end{subequations}

Lastly, as is mentioned in the main text, there is a sizeable difference between what is measured as the valley Zeeman g-factor of the trion $g_z$ and the expected contribution of the atomic orbital associated magnetic moment. This is most likely due to a significant shift of the experienced berry curvature $\Omega(\mathbf{k})$ and exchange energy of the trion\cite{yu2014dirac} which contributes to the valley g-factor\cite{srivastava2015valley}. This additional contribution is part of the measured $g_z$ from this experiment. It is known that there is an associated magnetic moment $\mu$ with the trion exchange energy

\begin{equation}
	\mu(\mathbf{k})=\frac{e}{2\hbar}\delta_{ex}\Omega(\mathbf{k})
\end{equation}

\noindent where $\delta_{ex}$ is the zero field exchange energy splitting between the singlet and triplet trion configurations which is $\delta_{ex}\sim\unit[4]{meV}$. The g factor associated with this magnetic moment is

\begin{equation}
	g_\Omega=\frac{m_e}{2\hbar^2}\delta_{ex}\Omega(\mathbf{k})
\end{equation}

\noindent The data presented in the main text suggests $g_\Omega$ is of magnitude $\sim4$, which from (13) suggests that $\Omega(\pm K)\sim \unit [10^4]{\text{\normalfont\AA}}$ at the $K$ $(K')$ valleys, in agreement with the expected value when modelling the trion with a massive Dirac Hamiltonian\cite{yu2014dirac}.